\documentclass[12pt]{article}
\usepackage[T1]{fontenc}
\usepackage{amsmath,amsfonts,amssymb}
\usepackage{graphics}
\usepackage{lscape}
\begin{document}
\newpage

\pagestyle{empty}
\vfill

\rightline{CERN-TH/2003-094}
\rightline{DSF-10/03}
\rightline{LAPTH-968/03}
\rightline{cond-mat/0304577}

\vfill

\begin{center}

{\Large \textbf{\textsf{Universality and Shannon entropy of codon usage}}}

\vspace{10mm}

{\large L. Frappat$^{ae}$, C. Minichini$^{b}$, A. Sciarrino$^{bd}$, P. 
Sorba$^{c}$}

\vspace{10mm}

\emph{$^a$ Laboratoire d'Annecy-le-Vieux de Physique Th{\'e}orique LAPTH}

\emph{CNRS, UMR 5108 associ{\'e}e {\`a} l'Universit{\'e} de Savoie}

\emph{BP 110, F-74941 Annecy-le-Vieux Cedex, France}

\vspace{5mm}

\emph{$^b$ Dipartimento di Scienze Fisiche, Universit{\`a} di Napoli
``Federico II''}

\emph{$^d$ I.N.F.N., Sezione di Napoli}

\emph{Complesso Universitario di Monte S. Angelo}

\emph{Via Cintia, I-80126 Naples, Italy}

\vspace{5mm}

\emph{$^c$ CERN, Theory Division, CH-1211 Geneva 23, Switzerland}

\vspace{5mm}

\emph{$^e$ Member of Institut Universitaire de France}

\vspace{7mm}

\end{center}

\vspace{7mm}

\begin{abstract}
The distribution functions of the codon usage probabilities, computed over 
all the available GenBank data, for 40 eukaryotic biological species and 5 
chloroplasts, do not follow a Zipf law, but are best fitted by the sum of a 
constant, an exponential and a linear function in the rank of usage. For 
mitochondriae the analysis is not conclusive. A quantum-mechanics-inspired 
model is proposed to describe the observed behaviour. These functions are 
characterized by parameters that strongly depend on the total $GC$ content 
of the \emph{coding regions} of biological species. It is predicted that 
the codon usage is the same in all exonic genes with the same $GC$ content. 
The Shannon entropy for codons, also strongly depending on the exonic $GC$ 
content, is computed.
\end{abstract}

\vfill
PACS number: 87.10.+e, 02.10.-v
\vfill

\vspace*{3mm} 

\hrule 

\vspace*{3mm} 

\noindent 
\emph{E-mail:} \texttt{frappat@lapp.in2p3.fr}, 
\texttt{minichini@na.infn.it}, \texttt{sciarrino@na.infn.it} \\
\texttt{sorba@lapp.in2p3.fr}, \texttt{sorba@cern.ch} (on leave of absence 
from LAPTH)

\newpage
\pagestyle{plain}
\setcounter{page}{1}
\baselineskip=16pt

\section{Introduction}

In the recent past, some interest has been shown in applying methods of 
statistical linguistics and information theory for the analysis of DNA 
sequences, in particular in investigating whether the frequency 
distribution of nucleotides or sequences of nucleotides follows Zipf's law 
\cite{Zipf}, and using the Shannon entropy to identify the redundancy or 
the bias of a nucleotides sequence. Let us recall that, at the end of the 
forties, Zipf remarked that, in natural languages and in many other 
domains, the distribution function follows an inverse power law, which can 
be described, denoting by rank $n=1$ the most used \textit{word}, by $n=2$ 
the next one and so on, and with $a>0$, by
\begin{equation}  
\label{eq:z}
f_{n} = \frac{f_{1}}{n^{a}} \;.
\end{equation}
In 1992, it was shown \cite{1992} that the distribution of the nucleotides 
in DNA follows a Zipf law of the type given by eq. (\ref{eq:z}). However, 
the opinions on this statement are divided, see \cite{92-93}. Soon after 
\cite{Mantegna94,Czirok95}, it was shown that noncoding sequences of DNA 
are more similar to natural languages than the coding ones, and the Shannon 
entropy has been used to quantify the \textit{redundancy} of word. This 
work also raised a debate in the literature \cite{1996}. Also, Zipf's law 
seems well adapted to represent the abundance of expressed genes, with an 
exponent $ a \approx 1 $, as can be seen in the recent work of ref. 
\cite{Furu02}. The origin of Zipf's law is thought to arise from stochastic 
processes \cite{CSV}, especially when they can be modelled as random walks 
in log scale \cite{KH}; for further analysis, see \cite{BA}. However, an 
analysis of the rank distribution for codons, performed in many genes for 
several biological species, led the authors of \cite{Som01} to fit 
experimental data with an exponential function. In particular, by 
considering separately different coding DNA sequences, they studied the 
relation between the parameter in the exponential, the frequency of the 
rank one and the length of the sequence for different genes. From this very 
short overview, it follows that the determination of the kind of law the 
codon rank distribution follows is extremely interesting, in investigations 
of the nature of the evolutionary process, which has acted upon the codon 
distribution, i.e. the eventual presence of a bias.

In the last few years, the number of available data for coding sequences 
has considerably increased, but apparently no analysis using the whole set 
of data has been performed. The aim of this paper is to perform such a 
study. As the result of our investigation, we will point out that the rank 
of codon usage probabilities follows a universal law, the frequency 
function of the rank-$n$ codon showing up as a sum of an exponential part 
and a linear part. A quantum-mechanics-inspired model is proposed, which 
provides for the form of the observed law and gives the correct sign in the 
different terms. Such a universal behaviour suggests the presence of 
general biases, one of which is identified with the total \emph{exonic} 
$GC$ content. Indeed, the values of the parameters appearing in the fitting 
expression are plotted versus the total percentage of exonic $GC$ content 
of the biological species and are reasonably well fitted by a parabola. 
Finally, from the obtained expression, we derive the theoretical prediction 
that the usage probability for \emph{rank-ordered} codons is the same in 
any genic region having the same exonic $GC$ content for any biological 
species.

We compute the Shannon entropy \cite{Sha} for amino-acids and find that its 
behaviour in function of the exonic $GC$ content is also a parabola, whose 
apex is around the value 0.50 of the $GC$ content.

\section{Codon usage probabilities distribution} 

Let us define the usage probability for the codon $XZN$ ($X, Z, N \in \{A,
C, G, U\}$) as
\begin{equation}  
\label{eq:1}
P(XZN) = \lim_{n_{tot} \to \infty} \;\;\; \frac{n_{XZN}}{N_{tot}} \;,
\end{equation}
where $n_{XZN}$ is the number of times the codon $XZN$ has been used in the 
analysed biosynthesis process for a given biological species and $N_{tot}$ 
is the total number of codons used in all considered processes. It follows 
that our analysis and predictions hold for biological species with 
sufficiently large statistics of codons. For each biological species, 
codons are ordered following the decreasing order of the values of their 
usage probabilities, i.e. codon number 1 corresponds to the highest value, 
codon number 2 is the next highest, and so on. We denote by $f(n)$ the 
probability $P(XZN)$ of finding $XZN$ is in the $n$-th position. Of course 
the same codon occupies in general two different positions in the rank 
distribution function for two different species. We plot $f(n)$ versus the 
rank and we determine that the best fit to the data can be reached by the 
sum of an exponential function, a linear function in the rank and a 
constant, i.e.
\begin{equation}
f(n) = \alpha \, e^{-\eta n} \, - \, \beta \, n \, + \, \gamma \;,
\label{eq:bf}
\end{equation}
where $0.0187 \leq \alpha \leq 0.0570$, $0.050 \leq \eta \leq 0.136$, $0.82 
\; 10^{-4} \leq \beta \leq 3.63 \; 10^{-4}$, $\gamma = 0,016$ are constant 
depending on the biological species. These four constants have to satisfy 
the normalization condition
\begin{equation}
\sum_{n} \, f(n) = 1 \;.
\label{eq:nor}
\end{equation}

\begin{center}
\begin{figure}
\includegraphics{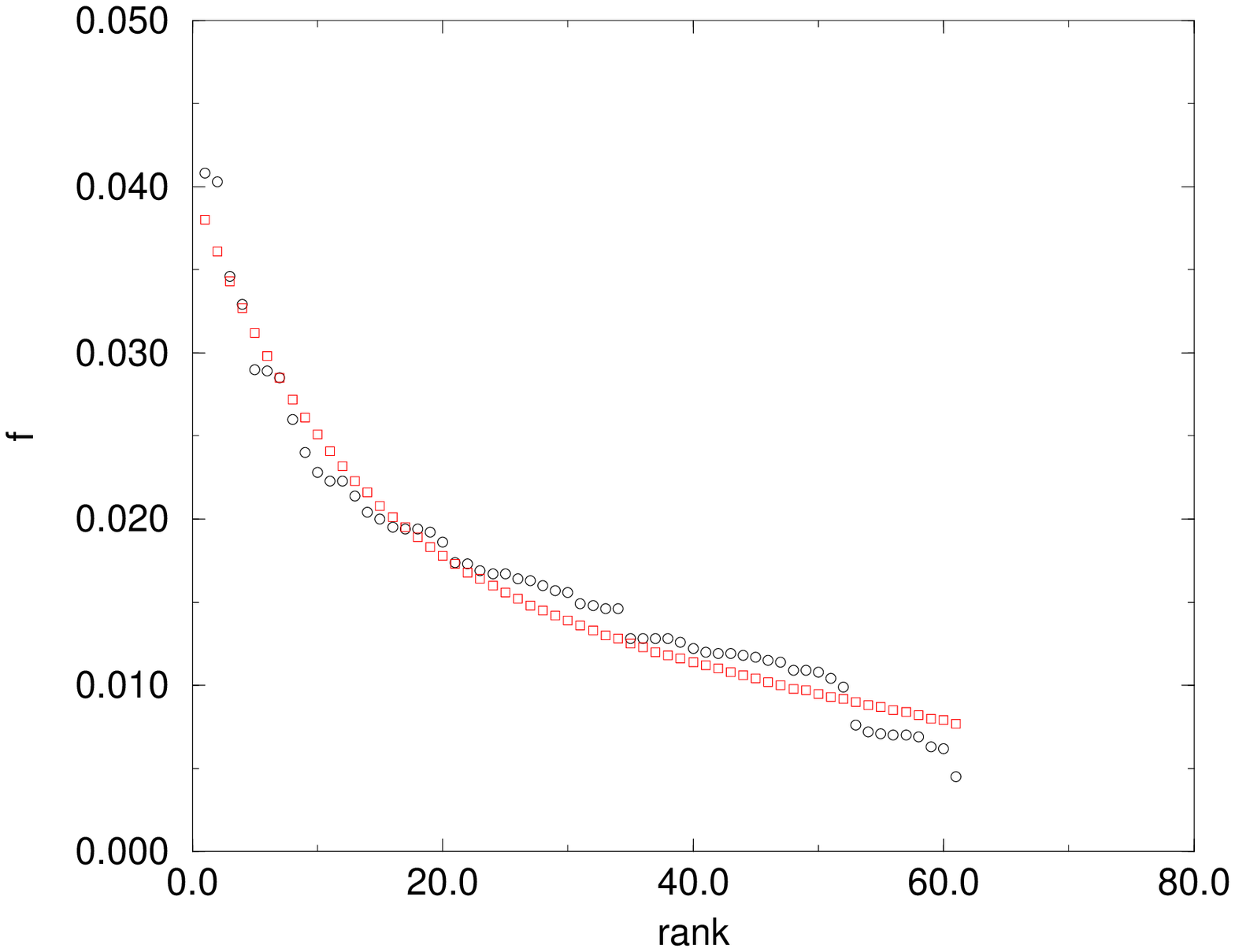}
\caption{Rank distribution of the codon usage probabilities for 
\textit{Homo sapiens}. Circles are experimental values, squares are fitted 
values.}
\label{fig:homo}
\end{figure}
\end{center}

\begin{center}
\begin{figure}
\includegraphics{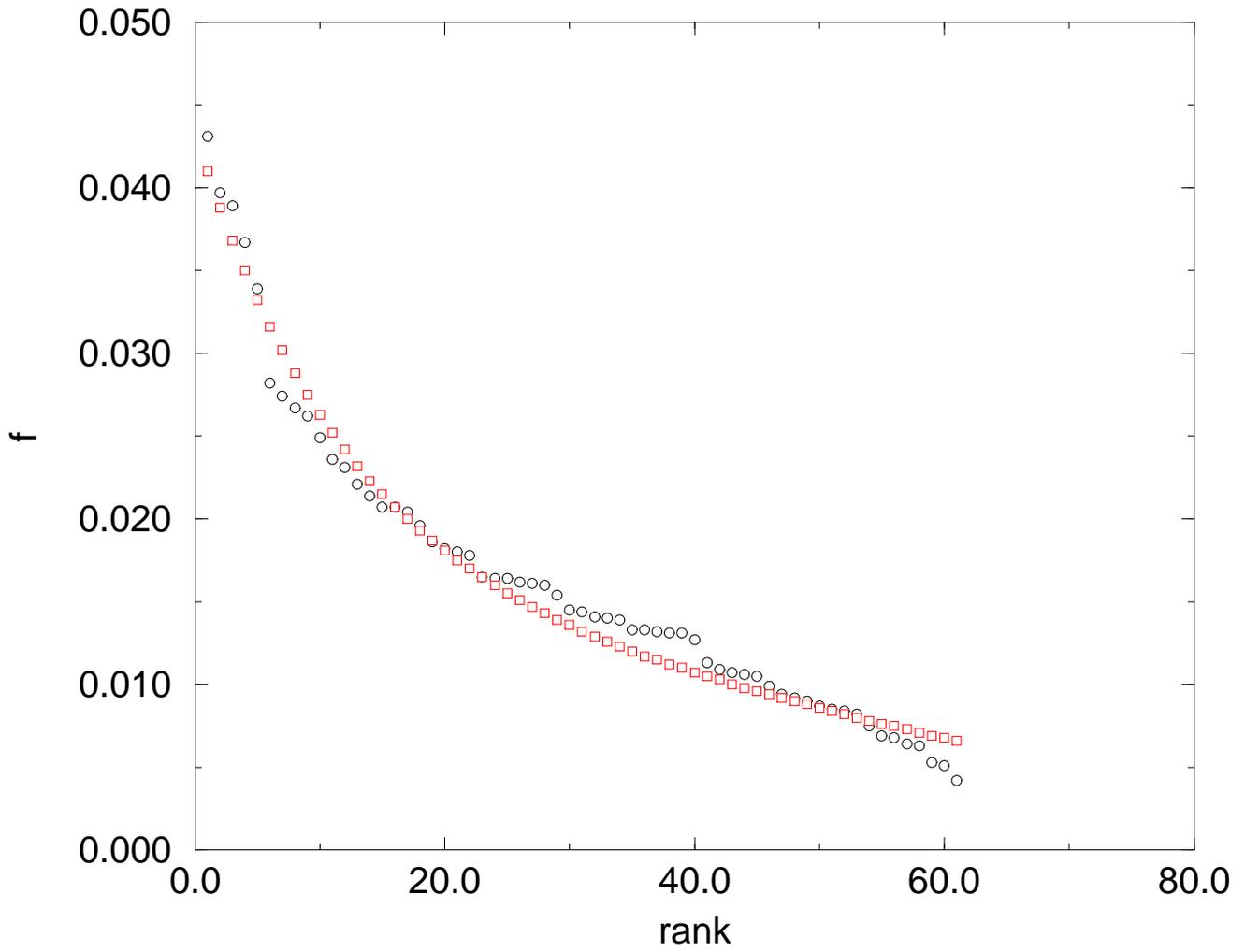}
\caption{Rank distribution of the codon usage probabilities for 
\textit{Drosophila melanogaster}. Circles are experimental values, squares 
are fitted values.}
\label{fig:dros}
\end{figure}
\end{center}

\begin{center}
\begin{figure}
\includegraphics{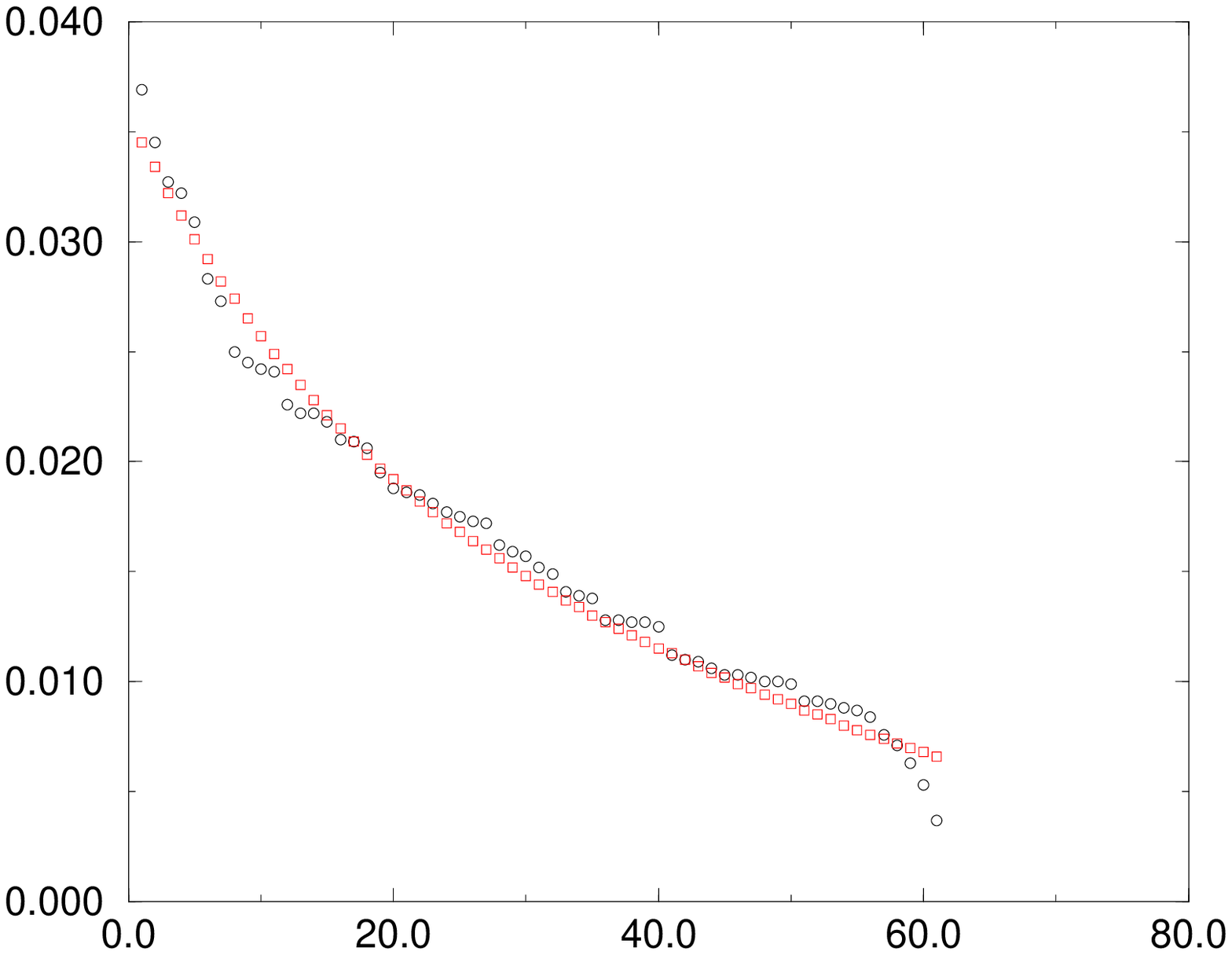}
\caption{Rank distribution of the codon usage probabilities for 
\textit{Arabidopsis thaliana}. Circles are experimental values, squares are 
fitted values.}
\label{fig:arab}
\end{figure}
\end{center}

\begin{center}
\begin{figure}
\includegraphics{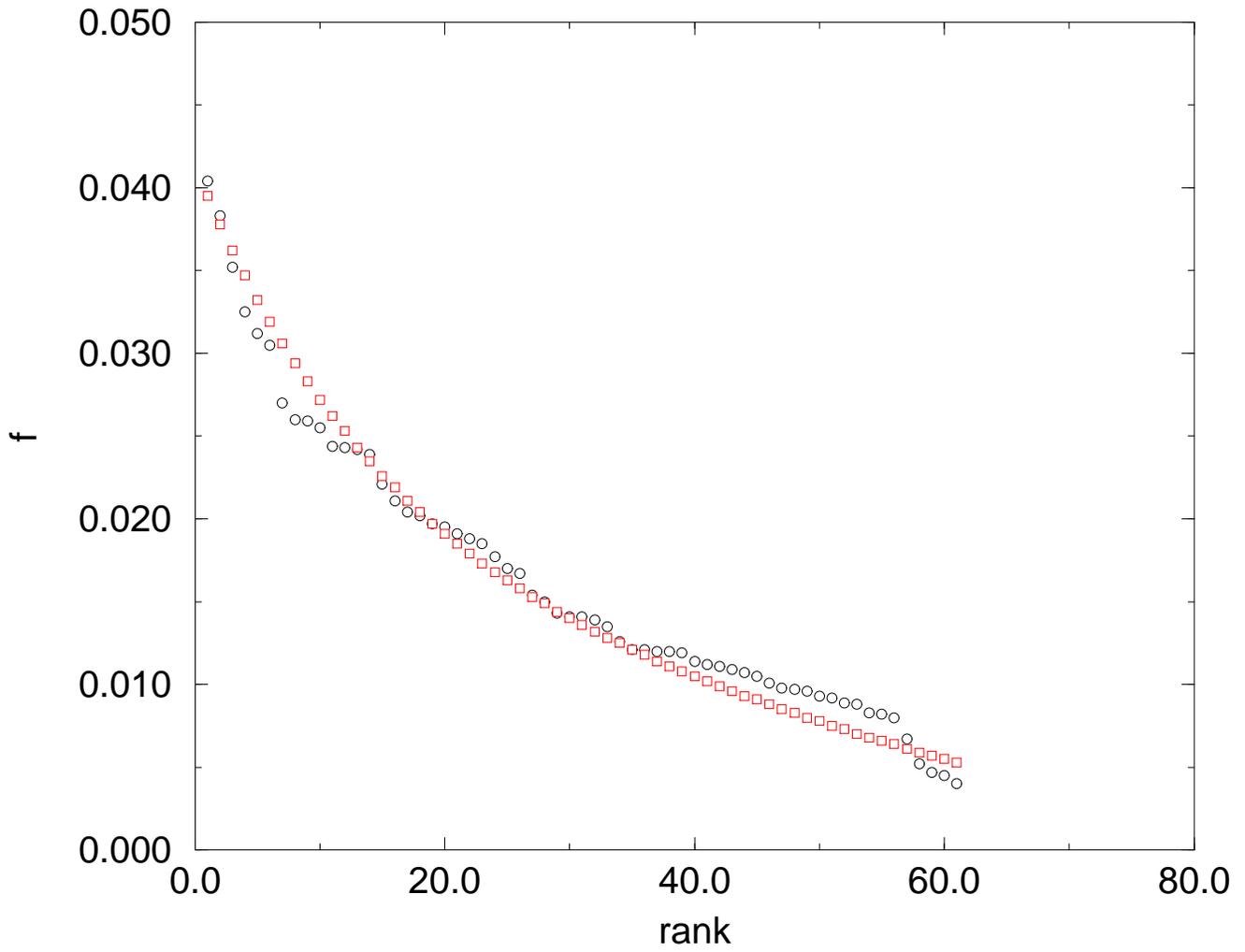}
\caption{Rank distribution of the codon usage probabilities for 
\textit{Escherichia coli}. Circles are experimental values, squares are 
fitted values.}
\label{fig:esch}
\end{figure}
\end{center}

\begin{center}
\begin{figure}
\includegraphics{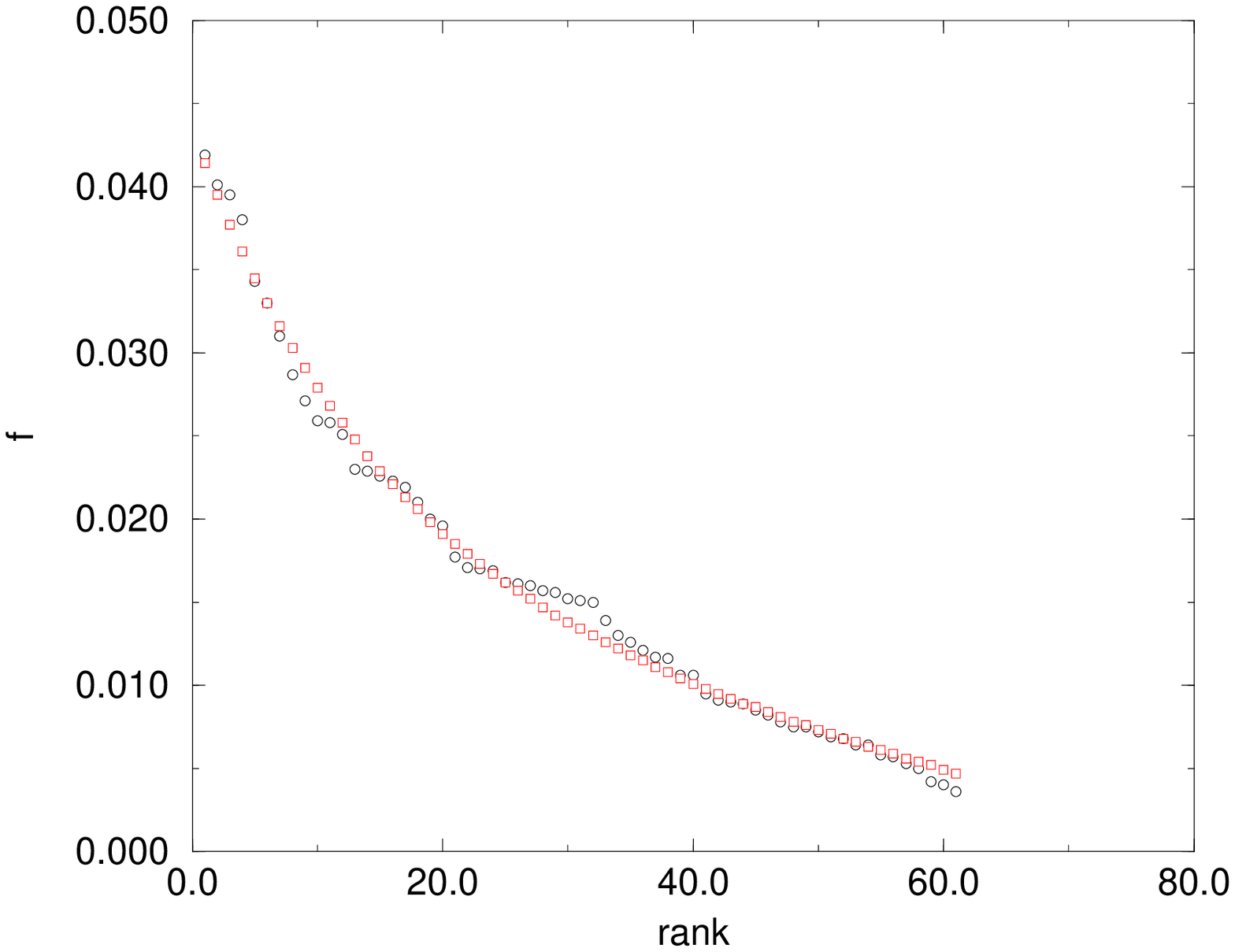}
\caption{Rank distribution of the codon usage probabilities for 
\textit{Chloroplast Arabidopsis thaliana}. Circles are experimental values, 
squares are fitted values.}
\label{fig:chl-arab}
\end{figure}
\end{center}

\begin{center}
\begin{figure}
\includegraphics{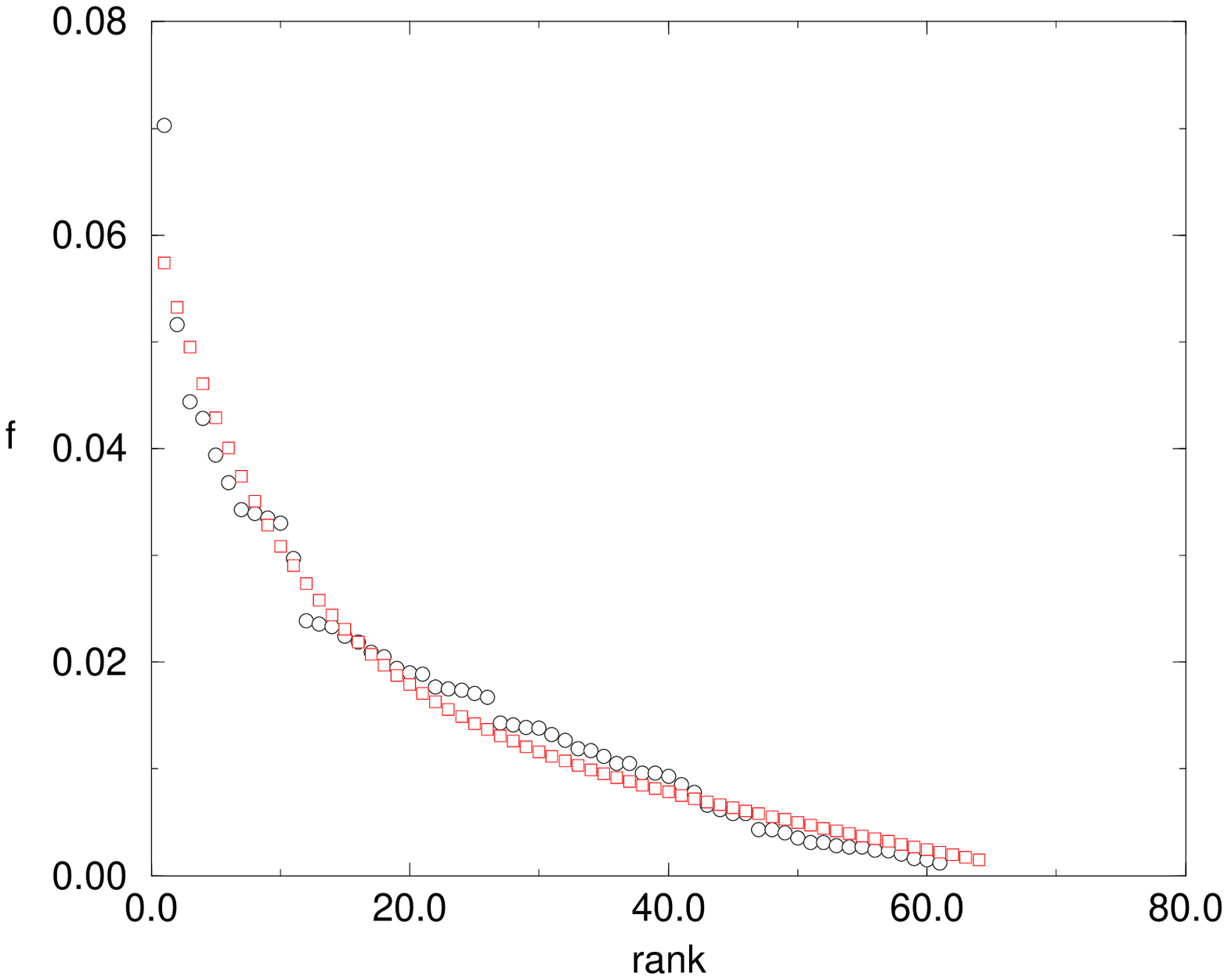}
\caption{Rank distribution of the codon usage probabilities for 
\textit{Mitochondrial Homo sapiens}. Circles are experimental values, 
squares are fitted values.}
\label{fig:mitohomo}
\end{figure}
\end{center}

\begin{center}
\begin{figure}
\includegraphics{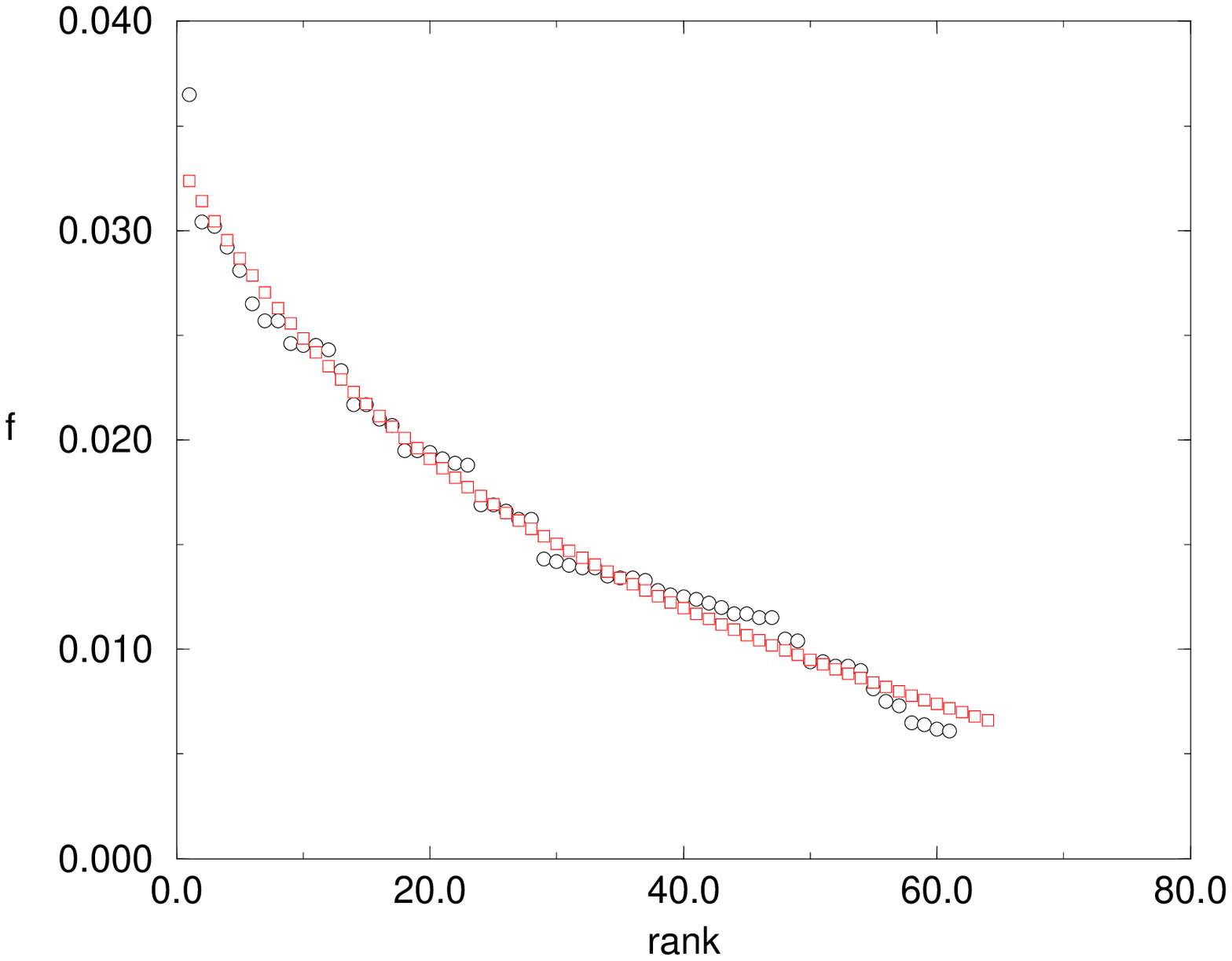}
\caption{Rank distribution of the codon usage probabilities for 
\textit{Mitochondrial Arabidopsis thaliana}. Circles are experimental values, 
squares are fitted values.}
\label{fig:mitoarab}
\end{figure}
\end{center}

In table \ref{table:data} we list the 40 biological species (6 vertebrates, 
4 plants, 3 invertebrates, 2 fungi and 25 bacteriae) -- with a number of 
codons ranging between 800\,000 and 20\,000\,000 in decreasing order (data 
from GenBank release 129.0 \cite{NGI}), whose codon usage has been fitted, 
specifying for each biological species the value of the parameters, 
computed by a best-fit procedure and the corresponding $\chi^{2}$. Here and 
in the following, the $\chi^2$ coefficient is defined by
\begin{equation}
\chi^2 = \sum_{i} \frac{(y_{i}-y(x_{i}))^2}{y(x_{i})} \;,
\label{eq:Rpearson}
\end{equation}
where $x_{i}$ are the experimental abscissae, $y_{i}$ the experimental 
values and $y(x_{i})$ the fitted ones. In some cases, $y(x_{i})$ takes 
vanishing or negative values for a few points, hence the $\chi^2$ is not 
reported. In figs. \ref{fig:homo}--\ref{fig:esch}, we report the plot of 
$f(n)$ as a function of $n$ for a few biological species (\textit{Homo 
sapiens}, \textit{Drosophila melanogaster}, \textit{Arabidopsis thaliana}, 
and \textit{Escherichia coli}). The plot has been cut to $n = 61$ to take 
into account the fact that in standard code there are three Stop codons (to 
end the biosynthesis process), whose function is very peculiar. For the 
same reason, the $\chi^2$ has been computed by taking into account the 61 
coding codons only. In table \ref{table:codons}, we report the type of the 
twenty-first most used codons of the observed rank distribution $f(n)$. A 
similar study, for a sample of twenty vertebrates with a codon statistics 
larger than 100\,000, reveals that, for almost all biological species, the 
four most used codons are $GAG$, $CUG$, $AAG$ and $CAG$. All these codons 
have a $G$ nucleotide in third position and three of them encode doublets. 
An analysis performed on the chloroplast codon usage for a sample of five 
plants gives the same result for the rank distribution $f(n)$, see table 
\ref{table:datachl} and fig. \ref{fig:chl-arab} (\textit{Chloroplast 
Arabidopsis thaliana}). We also report, in table \ref{table:datamito}, the 
values of the parameters and the $\chi^2$ for a sample of nine 
mitochondriae with a codon statistics larger than 15\,000. The fits for 
\textit{Homo sapiens} and \textit{Arabidopsis thaliana} are presented in 
figs. \ref{fig:mitohomo} and \ref{fig:mitoarab}. We point out, however, 
that for mitochondriae the codon usage frequencies distribution for several 
species (e.g. \textit{Arabidopsis thaliana} or \textit{Drosophila 
melanogaster}) is ill fitted by eq. (\ref{eq:bf}). It may be an indication 
that mitochondriae do not follow the universal law (\ref{eq:bf}). Note that 
the mitochondrial codes have a few differences with the eukaryotic code and 
vary slightly between species, see e.g. \cite{LAPTH787}. In these cases, 
the $\chi^2$ has been computed over the corresponding coding codons. The 
value of the constant $\gamma$ is approximately equal to $1/61 = 0.0164$ or 
to $1/64 = 0.0156$, i.e. the value of the codon usage probability in the 
case of uniform and not biased codon distribution. Therefore the other two 
terms in eq. (\ref{eq:bf}) can be viewed as the effect of the bias 
mechanism. The appearance of the linear term is more intriguing. Let us 
remark that in \cite{Som01}, where an exponential function is used to fit 
the rank of usage in genes (not the rank of usage probability), the linear 
term has not been observed, as its contribution becomes noticeable for 
approximately $n \ge 20$. Owing to the analysis of genes (with at most a 
few hundred codons), the fits in that paper end before this value of the 
rank. It is believed that the main causes of codon usage bias are the 
translational efficiency, the selection pressure and the spontaneous 
mutations. From the smallness of parameter $\beta$ in (\ref{eq:bf}), it is 
tempting to identify the latter as the consequence of mutation effect and 
the first term in (\ref{eq:bf}) as the effect of selection pressure, i.e. 
the interaction with the environment.

\medskip

Let us try to build up a simple model to explain the universal structure of 
eq. (\ref{eq:bf}). In a physical quantum system, with discrete energy 
levels, the occupation number of the $i$-th energetic level is proportional 
to $ \exp(-E_{i}/kT)$, where $ E_{i}$ is the energy of the level, $T$ is 
identified with the temperature and $k$ is the Boltzmann constant. Inspired 
by this analogy, let us assume that the rank of the probability 
distribution is a function of some kind of \textit{energy}. In this case 
the codon usage probability is, for any codon, equal to $f(n) = \exp(-a 
E_{n})$, where $a$ is a positive constant. We assume that the system 
satisfies periodic conditions in the $y$ and $z$ directions, with a very 
small spacing in the $z$ direction, so that the excited modes along this 
direction can be neglected, and that the interaction with the environment 
can be modelled as the switching on of a constant uniform \textit{magnetic 
field} in the direction of the $z$ axis. As a consequence, the Landau 
levels appear, the \textit{energetic levels} now being in natural units
\begin{equation}
\label{eq:en} 
E_{n} -  E_{0} =  \omega (n + 1/2) \;,
\end{equation}
where $\omega$ is the so-called cyclotron frequency. The wave function is
\begin{equation}
\label{eq:wf} 
\psi(x,y,z) = \Phi_n(x - lx_0) \, e^{ik_y y} \, e^{ik_z z} \;,
\end{equation}
where
\begin{equation}
k_y = \frac{2 \pi l}{L_y} \;, \;\;\;\;\;\;\; k_z = \frac{2 \pi}{L_z} \;, 
\;\;\;\;\;\;\; l \in {\bf Z}, l \ne 0 \;,
\end{equation}
and $\Phi_n(x - lx_0)$ is the wave function of the $n$-th energy level of a 
one-dimensional harmonic oscillator shifted by $lx_0$ ($x_0 = 2 \pi/q 
BL_y$). There is a degeneracy due to the arbitrariness of $k_y$. A quantun 
charged physical system in an excited state spontaneously falls down to a 
lower state, the equilibrium configuration being the system in the 
fundamental state. In the case of codons, the equilibrium condition is the 
configuration where they occupy different ``energy'' levels, the situation 
being the analogue of a quantum system in the presence of an external 
pumping-back source. In order to model this situation in a simple way, we 
assume that the equilibrium probability distribution is a function $f(n) - 
f(n)_{B=0} = F(E_n, \gamma_n)$, where $\gamma_n$ is the ``spontaneous 
emission coefficient'' of the $n$-th level, which is taken to be very small 
because of the assumed equilibrium condition. We can make a first-order 
development in $\gamma_{n}$:
\begin{equation}
f(n) - f(n)_{B=0} = F(E_n, \gamma_n = 0) + \frac{dF}{d \gamma_n} \;,
\end{equation}
where $dF/d \gamma_n$ is negative, the occupation probability decreasing 
with the increasing emission. We compute $\gamma_n$ in the standard dipole 
approximation \cite{Sakurai} between the $n$-th and the $k$-th levels 
($k<n$). In the case of our model, we get
\begin{equation}
\gamma_n \varpropto n \, \delta_{k,n-1} \;.
\end{equation}
So, replacing $F(E_n,\gamma_n = 0)$ by the Maxwell--Boltzmann expression, 
we obtain eq. (\ref{eq:bf}). The values of the parameters depend on the 
biological species, as the selection pressure (in the model, the value of 
the \textit{charge} and of the \textit{magnetic field}) is different for 
different biological species. Even if our simple model is able to explain 
the universal behaviour and the right sign of the linear term, one should 
rather consider it as a toy model. However, the fact that one can build a 
simple mathematical scenario, able to reproduce the observed distribution, 
provides us with an indication of the existence of some strong 
physico-chemical constraints that add to the random effects.

Since it is well known that the $GC$ content plays a strong role in the 
evolutionary process, we expect the parameters to depend on the total $GC$ 
content of the genes region (here the total exonic $GC$ content), which is 
indeed correlated with the evolution of the system (see \cite{KFL} and 
references therein). We have investigated this dependence and report, in 
table \ref{table:alphaGC}, the fits of $\alpha$ and $\beta$ to the total 
exonic $GC$ content $Y_{GC}$ of the biological species. One finds that the 
values of $\alpha$ and $\beta$ are well fitted by polynomial functions 
(with $0 \le Y_{GC} \le 100$ in percentage):
\begin{eqnarray}
\label{eq:fitalpha}
\alpha &=& 0.21145 - 0.00776 \, Y_{GC} + 7.92 \; 10^{-5} \, Y_{GC}^2 \;, 
\qquad \chi^2 = 0.0262 \;, \\
\label{eq:fitbeta}
10^{2} \, \beta &=& 0.10096 - 0.00345 \, Y_{GC} + 3.50 \; 10^{-5} \, 
Y_{GC}^2 \;, \qquad \chi^2 = 0.0170 \;.
\end{eqnarray}
The two parameters $\alpha$ and $\beta$ appear to be correlated. Indeed the 
plot representing $\beta$ as a function of $\alpha$ is satisfactorily 
fitted by a regression line (see fig. \ref{table:entropyGC}):
\begin{eqnarray}
\label{eq:fitba}
10^{2} \, \beta &=& 0.00851 + 0.375 \, \alpha \;, \qquad \chi^2 = 0.0218 
\;.
\end{eqnarray}
The value of the $\eta$ parameter is largely uncorrelated with the total 
exonic $GC$ content. Let us recall that, however, $\eta$ is a function of 
$\alpha$ and $\beta$ due to the normalization condition of eq. 
(\ref{eq:nor}). Indeed we have (assuming $ e^{-65 \eta} \approx 0$) 
\footnote{Note that the result is almost unchanged if the data are 
normalized on the 61 coding codons.}
\begin{equation}
\label{eq:eta}
1 = \frac{\alpha \, e^{-\eta}}{1 - e^{-\eta}} + 2080 \beta + 64 \gamma \;.
\end{equation}
Using the fits for $\alpha$ and $\beta$, we can write the probability 
distribution function for any biological species, whose total $GC$ content 
in per cent in the exonic regions is $Y_{GC}$, as
\begin{equation}
\label{eq:thf}
f(n) = (\alpha_0 + \alpha_1 Y_{GC} + \alpha_2 Y_{GC}^2) e^{-\eta n} - 
n(\beta_0 + \beta_1 Y_{GC} + \beta_2 Y_{GC}^2) + \gamma \;,
\end{equation}
where $\eta$ is obtained by solving eq. (\ref{eq:eta}). Of course we are 
not able to predict which codon occupies the $n$-th rank. Finally, let us 
remark that the total exonic $GC$ content $Y_{GC}$ has to satisfy the 
consistency condition
\begin{equation}
\label{eq:YGC}
Y_{GC} =  \frac{1}{3} \, \sum_{i \in I} \, d_i f(i) \;,
\end{equation}
where the sum is over the set $I$ of integers to which the 56 codons 
containing $G$ and/or $C$ nucleotides belong and $d_i$ is the 
multiplicity of these nucleotides inside the $i$-th codon.

\section{Amino-acids rank distribution}

It is natural to wonder if some kind of universality is also present in the 
rank distribution of amino-acids. From the available data for codon usage, 
we can immediately compute (using the eukaryotic code) the frequency of 
appearance of any amino-acid $F(n)$ ($1 \le n \le 20$) in the whole set of 
coding sequences. The calculated values as a function of the rank are 
satisfactorily fitted by a straight line,
\begin{equation}
\label{eq:FAA}
F(n) = F_{0} - Bn \;.
\end{equation}
The parameters $F_{0}$, $B$ and the corresponding $\chi^2$ for the fits are 
reported in table \ref{table:amino}. A better fit can be obtained in 
general by using a third-degree polynomial; however, the range of the four 
parameters for this fit is larger than the range of the 2-parameter fit. 
For a few biological species, we give below the parameters for the two 
fits, see also the figures of table \ref{table:lincub}. The plots of the 
linear fits for few biological species are given in table 
\ref{table:figures}. Note that the 21st point is just the contribution of 
the Stop codons, which of course has not been taken into account for the 
fits. One can remark that the most frequent amino-acid is always above the 
line. This can be easily understood in the light of eq. (\ref{eq:bf}). 
Indeed, the most frequent amino-acids get, in general, a contribution of 
the exponential term of (\ref{eq:bf}) with a low value of $n$.

\begin{table}[htbp]
\centering
\begin{tabular}{|l|ll|l|}
\hline
Species & \multicolumn{2}{|l|}{linear/cubic fits} & $\chi^2$ \\
\hline
Homo sapiens & lin. & $f = 0.087 - 0.0036 n$ & 0.0072 \\
 & cub. & \!\! $f = 0.099 - 0.0088 n + 57 \, 10^{-5} n^2 - 1.7 \, 10^{-5} 
 n^3$ & 0.0055 \\
\hline
Arabidopsis thaliana & lin. & $f = 0.088 - 0.0036 n$ & 0.0068 \\
 & cub. & $f = 0.099 - 0.0090 n + 62 \, 10^{-5} n^2 - 1.95 \, 10^{-5} n^3$ 
 & 0.0049 \\
\hline
Drosophila melanogaster & lin. & $f = 0.087 - 0.0036 n$ & 0.0125 \\
 & cub. & $f = 0.097 - 0.0096 n + 76 \, 10^{-5} n^2 - 2.5 \, 10^{-5} n^3$ & 
 0.0042 \\
\hline
Escherichia coli & lin. & $f = 0.090 - 0.0039 n$ & 0.0115 \\
 & cub. & $f = 0.112 - 0.0136 n + 105 \, 10^{-5} n^2 - 3.1 \, 10^{-5} n^3$ 
 & 0.0067 \\
\hline
\end{tabular}
\end{table}

Of course, the frequency of an amino-acid is given by the sum of the 
frequencies of its encoding codons given by (\ref{eq:bf}). If the ranks of 
the encoding codons were completely random, we do not expect that their sum 
should take equally spaced values, as is the case in a regression line. 
Therefore, we can infer, for the biological species whose amino-acids 
frequency is very well fitted by a line, the existence of some functional 
constraints on the codon usage.

We report in table \ref{table:acidamines} the distribution of the 
amino-acids for the different biological species. 

However, the behaviour predicted by eq. (\ref{eq:bf}) fits the experimental 
data very well, while the shape of the distribution of amino-acids seems 
more sensible to the biological species. In fact, one can remark on many 
plots of the amino-acid distributions (see e.g. table \ref{table:figures}), 
the existence of one or two plateaux, for which we do not have any 
explanation yet.

\section{Consequences of probability distribution}

We now derive a few consequences of eq. (\ref{eq:bf}). In the following, we 
denote by $y$ the \emph{local} exonic $GC$ content (i.e. for coding 
sequences of genes) for a given biological species. Let us assume that the 
exonic $GC$ content of a biological species is essentially comprised in the 
interval $y_1 - y_0 = \Delta$ (e.g. for \textit{Homo sapiens} $y_0 = 35 \%$ 
and $y_1 = 70 \%$). We can write
\begin{equation}
\label{eq:finteg}
f(n) = \frac{1}{\Delta} \int_{y_0}^{y_1} \, f(y,n) dy \;.
\end{equation}
Since the l.h.s. of the above equation has the form given by eq. 
(\ref{eq:bf}) for any $n$ and for any biological species, if we do not want 
to invoke some ``fine tuning'' in the integrand function $f(y,n)$, we have 
to assume that
\begin{equation}
\label{eq:thfl}
f(y,n) = a(y) e^{-\eta n} - b(y) n + \gamma
\end{equation}
with the condition
\begin{equation}
\alpha = \frac{1}{\Delta} \, \int_{y_0}^{y_1} \, a(y) dy \;, \;\;\;\;\; 
\beta = \frac{1}{\Delta} \, \int_{y_0}^{y_1} \, b(y) dy \;.
\end{equation} 
As a consequence, we predict that the codon usage probability is the same 
for any codon in any exonic genic region with the same $GC$ content. The 
form of the $a(y)$ and $b(y)$ functions is yet undetermined. \\
For \textit{Homo sapiens}, we remark that the total exonic $GC$ content 
$Y_{GC}$ is, in a very good approximation, equal to the mean value of the 
interval $[y_{0},y_{1}]$. Therefore, inserting (\ref{eq:finteg}) and 
(\ref{eq:thfl}) into (\ref{eq:YGC}), we derive that the functions $a(y)$ 
and $b(y)$ have to be \emph{linear} functions of $y$. This theoretical 
derivation is in accordance with the conclusions of Zeeberg \cite{Zee} 
obtained by an analysis for 7357 genes. On a quantitative level, using the 
numerical linear fits of Zeeberg, we find a very good agreement with our 
calculations. Note that this result is not in contradiction with eq. 
(\ref{eq:thf}), since the previous analysis is valid for the fixed value of 
the exonic $GC$ content for \textit{Homo sapiens}. For bacteriae, the range 
of variation $\Delta$ of the local exonic $GC$ content is very small. 
Therefore we expect the functions $a(y)$ and $b(y)$ to have the same shape 
as the functions $\alpha$ and $\beta$ given in eqs. (\ref{eq:fitalpha}) and 
(\ref{eq:fitbeta}). Hence the functions $\alpha$ and $\beta$ depend on the 
biological species. 

We compute the Shannon entropy,  given by
\begin{equation}
S = - \sum_n \, f(n) \, \log_2 f(n)
\end{equation}
for the codons of a biological species and plot it versus the total exonic 
$GC$ content; see figure in table \ref{table:entropyGC}. The Shannon 
entropy is rather well fitted by a parabola:
\begin{equation}
\label{eq:fitS}
S = 2.2186 + 0.144 \, Y_{GC} - 0.00146 \, Y_{GC}^2 \;, \qquad \chi^2 = 
0.0315 \;.
\end{equation}
Note that the parabola has its apex for $y \approx 0.50$, which is expected 
for the behaviour of the Shannon entropy for two variables (here $GC$ and 
its complementary $AU$).

The same behaviour have been put in evidence by analogous computations made 
by Zeeberg \cite{Zee} for \textit{Homo sapiens}. So it seems that the 
entropy in the gene coding sequences and the total exonic region entropy in 
function of the exonic $GC$ content show the same pattern.

\bigskip

In conclusion, the distribution of the experimental codon probabilities for 
a large total exonic region of several biological species has been very 
well fitted by the law of eq. (\ref{eq:bf}). The spectrum of the 
distribution is universal, but the codon, which occupies a fixed level, 
depends on the biological species. Indeed, a more detailed analysis shows 
that, for close biological species, e.g. vertebrates, a fixed codon 
occupies almost the same position in $f(n)$, while for distant biological 
species the codons occupy very different positions in the rank 
distribution. We have also derived that the codon frequency for any genic 
region is, for fixed biological species and fixed $GC$ content, the same. 
Entropy analysis has shown that the behaviour observed in the genes with 
different $GC$ content for the same biological species is very similar to 
the one shown by the total exonic region with different $GC$ content for 
the different biological species.

\bigskip

\textbf{Acknowledgements:} The authors would like to thank A. Arneodo for 
fruitful discussions. A.S. is indebted to the Universit{\'e} de Savoie for 
financial support and LAPTH for its kind hospitality.

\begin{table}
\centering
\caption{Values of the best-fit parameters, eq. (\ref{eq:bf}), for the 
sample of biological species. Types: vrt = vertebrates (6), inv = 
invertebrates (3), pln = plants (4), fng = fungi (2), bct = bacteriae (25).
\label{table:data}}
\begin{tabular}{|c|l|c|c|c|c|c|}
\hline
Type & Species & $GC$ content in \% & $\alpha$ & $\eta$ & $10^{4}\,\beta$ & 
$\chi^2$ \\
\hline
\hline
vrt & Homo sapiens & 52.58 & 0.0214 & 0.073 & 1.65 & 0.0126 \\
\hline
pln & Arabidopsis thaliana & 44.55 & 0.0185 & 0.056 & 1.68 & 0.0051 \\
\hline
inv & Drosophila melanogaster & 54.03 & 0.0247 & 0.081 & 1.67 & 0.0089 \\
\hline
inv & Caenorhabditis elegans & 42.79 & 0.0216 & 0.064 & 1.79 & 0.0063 \\
\hline
vrt & Mus musculus & 52.38 & 0.0208 & 0.071 & 1.57 & 0.0112 \\
\hline
fng & Saccharomyces cervisiae & 39.69 & 0.0246 & 0.069 & 1.91 & 0.0127 \\
\hline
bct & Escherichia coli & 50.52 & 0.0233 & 0.065 & 1.91 & 0.0112 \\
\hline
vrt & Rattus norvegicus & 52.87 & 0.0222 & 0.073 & 1.63 & 0.0083 \\
\hline
pln & Oryza sativa japonica & 55.84 & 0.0179 & 0.073 & 1.63 & 0.0211 \\
\hline
fng & Schizosaccharomyces pombe & 39.80 & 0.0255 & 0.068 & 1.98 & 0.0036 \\
\hline
bct & Bacillus subtilis & 44.32 & 0.0259 & 0.084 & 1.71 & 0.0241 \\
\hline
bct & Pseudomonas aeruginosa & 65.70 & 0.0538 & 0.107 & 2.76 & 0.0191 \\
\hline
bct & Mesorhizobium loti & 63.05 & 0.0416 & 0.093 & 2.44 & 0.0093 \\
\hline
bct & Streptomyces coelicolor A3 & 72.41 & 0.0567 & 0.098 & 3.14 & 0.0456 
\\
\hline
bct & Sinorhizobium meliloti & 62.71 & 0.0359 & 0.076 & 2.54 & 0.0067 \\
\hline
bct & Nostoc sp. PCC 7120 & 42.36 & 0.0288 & 0.098 & 1.63 & 0.0140 \\
\hline
pln & Oryza sativa & 54.63 & 0.0173 & 0.062 & 1.59 & 0.0135 \\
\hline
bct & Agrobacterium tumefaciens str. C58 & 59.74 & 0.0308 & 0.067 & 2.43 & 
0.0100 \\
\hline
bct & Ralstonia solanacearum & 67.57 & 0.0543 & 0.105 & 2.87 & 0.0149 \\
\hline
bct & Yersinia pestis & 48.97 & 0.0179 & 0.040 & 2.17 & 0.0066 \\
\hline
bct & Methanosarcina acetivorans str. C24 & 45.17 & 0.0228 & 0.068 & 1.81 & 
0.0214 \\
\hline
bct & Vibrio cholerae & 47.35 & 0.0203 & 0.052 & 2.02 & 0.0100 \\
\hline
bct & Escherichia coli K12 & 51.83 & 0.0250 & 0.065 & 2.05 & 0.0117 \\
\hline
bct & Mycobacterium tubercolosis CDC1551 & 65.77 & 0.0401 & 0.094 & 2.35 & 
0.0105 \\
\hline
bct & Mycobacterium tubercolosis H37Rv & 65.90 & 0.0414 & 0.097 & 2.29 & 
0.0109 \\
\hline
bct & Bacillus halodurans & 44.32 & 0.0263 & 0.100 & 1.27 & 0.0233 \\
\hline
bct & Clostridium acetobutylicum & 31.59 & 0.0434 & 0.087 & 2.76 & -- \\
\hline
bct & Caulobacter crescentus CB15 & 67.68 & 0.0570 & 0.113 & 2.86 & 0.0087 
\\
\hline
vrt & Gallus gallus & 52.11 & 0.0239 & 0.095 & 1.17 & 0.0129 \\
\hline
bct & Synechocystis sp. PCC6803 & 48.56 & 0.0260 & 0.083 & 1.49 & 0.0140 \\
\hline
bct & Sulfolobulus solfataricus & 36.47 & 0.0290 & 0.066 & 2.26 & 0.0099 \\
\hline
bct & Mycobacterium leprae & 59.90 & 0.0252 & 0.071 & 1.80 & 0.0065 \\
\hline
bct & Brucella melitensis & 58.25 & 0.0294 & 0.067 & 2.25 & 0.0121 \\
\hline
bct & Deinococcus radiodurans & 67.24 & 0.0481 & 0.098 & 2.76 & 0.0113 \\
\hline
vrt & Xenopus laevis & 47.33 & 0.0193 & 0.084 & 0.92 & 0.0268 \\
\hline
bct & Listeria monocytogenens & 38.39 & 0.0437 & 0.136 & 1.64 & 0.0267 \\
\hline
pln & Neurospora crassa & 56.17 & 0.0241 & 0.086 & 1.31 & 0.0166 \\
\hline
bct & Clostridium perfrigens & 29.47 & 0.0510 & 0.092 & 3.11 & -- \\
\hline
inv & Leishmania major & 63.36 & 0.0294 & 0.069 & 2.21 & 0.0050 \\
\hline
vrt & Bos taurus & 53.05 & 0.0240 & 0.089 & 1.27 & 0.0126 \\
\hline
\end{tabular}
\end{table}

\clearpage

\begin{table}[htbp]
\centering
\caption{Values of the best-fit parameters, eq. (\ref{eq:bf}), for the 
sample of chloroplasts.
\label{table:datachl}}
\begin{tabular}{|l|c|c|c|c|c|c|}
\hline
Species & $GC$ content in \% & $\alpha$ & $\eta$ & $10^{4}\,\beta$ & 
$\chi^2$ \\
\hline
\hline
Arabidopsis thaliana & 38.37 & 0.0254 & 0.067 & 1.95 & 0.0030 \\
\hline
Chaetosphaeridium globosum & 30.29 & 0.0515 & 0.110 & 2.59 & 0.0174 \\
\hline
Chlorella vulgaris & 34.63 & 0.0513 & 0.114 & 2.04 & 0.0093 \\
\hline
Cyanidium caldarium & 33.31 & 0.0379 & 0.092 & 2.24 & 0.0103 \\
\hline
Guillardia theta & 33.20 & 0.0452 & 0.103 & 2.20 & 0.0089 \\
\hline
\end{tabular}
\end{table}

\begin{table}[htbp]
\centering
\caption{Values of the best-fit parameters, eq. (\ref{eq:bf}), for the 
sample of mitochondriae.
\label{table:datamito}}
\begin{tabular}{|c|l|c|c|c|c|c|c|}
\hline
Type & Species & $GC$ content in \% & $\alpha$ & $\eta$ & $10^{4}\,\beta$ & 
$\chi^2$ \\
\hline
\hline
vrt & Homo sapiens & 44.99 & 0.0414 & 0.099 & 2.31 & 0.0207 \\
\hline
pln & Arabidopsis thaliana & 44.18 & 0.0136 & 0.049 & 1.39 & 0.0589 \\
\hline
vrt & Mus musculus & 37.23 & 0.0455 & 0.104 & 2.44 & 0.0226 \\
\hline
fng & Saccharomyces cerevisiae & 24.17 & 0.0879 & 0.198 & 2.66 & 0.0611 \\
\hline
inv & Physarum polycephalum & 25.69 & 0.0624 & 0.128 & 2.70 & 0.0262 \\
\hline
pln & Pylaiella littoralis & 37.06 & 0.0336 & 0.108 & 1.72 & 0.0112 \\
\hline
pln & Neurospora crassa & 33.20 & 0.0388 & 0.101 & 2.14 & 0.0225 \\
\hline
vrt & Bos taurus & 39.73 & 0.0422 & 0.106 & 2.25 & 0.0430 \\
\hline
vrt & Sus scrofa & 40.52 & 0.0497 & 0.112 & 2.51 & 0.0372 \\
\hline
\end{tabular}
\end{table}
\clearpage

\clearpage

\clearpage

\clearpage

\clearpage

\clearpage

\clearpage

\clearpage

\landscape
\begin{table}[htbp]
\centering
\label{table:alphaGC}
\begin{tabular}{cc}
& \\
\includegraphics{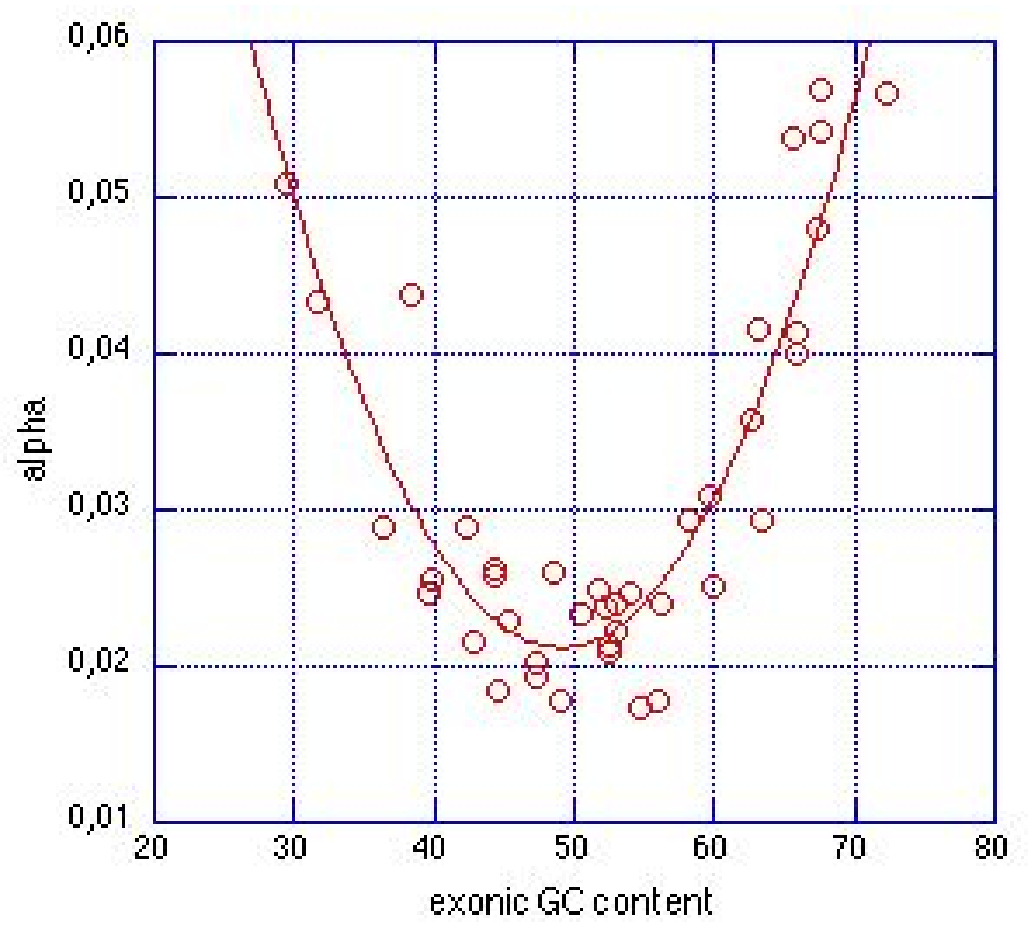} & \includegraphics{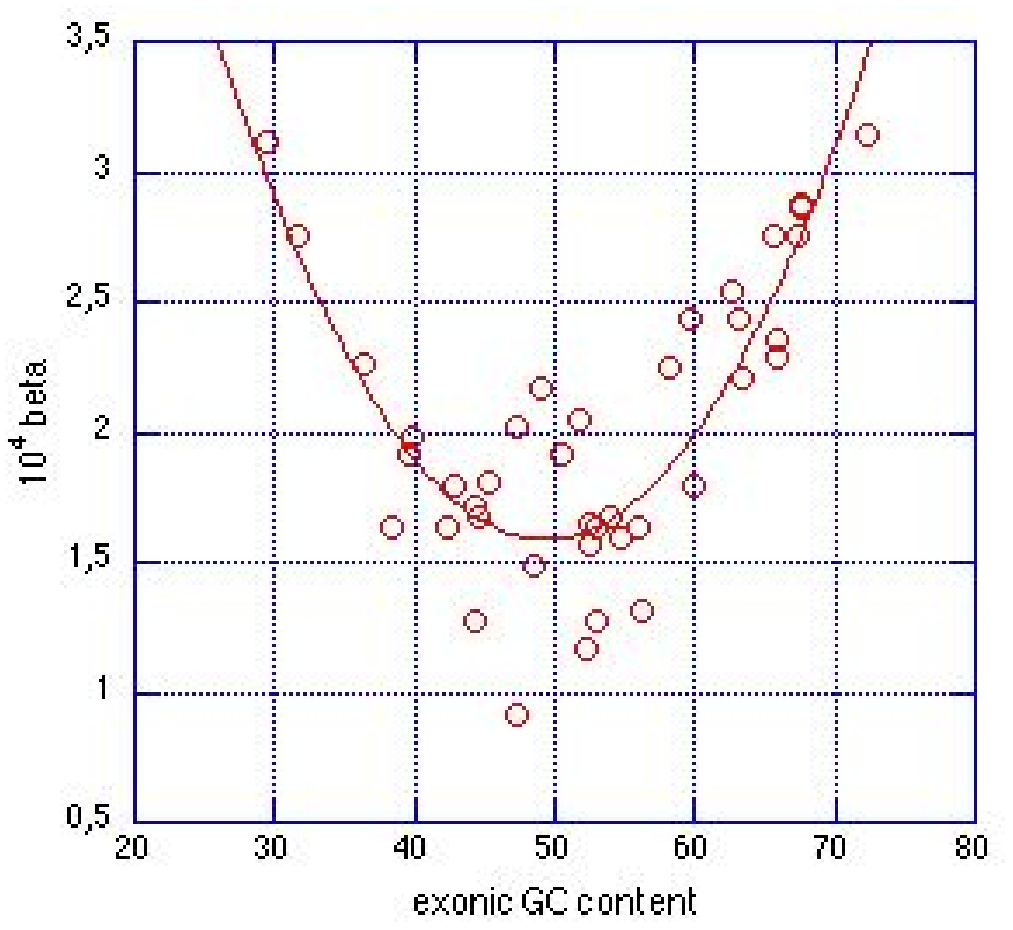} \\
& \\
$\alpha$ vs. the exonic $GC$ content & $\beta$ vs. the exonic $GC$ content 
\\
\end{tabular}
\caption{Fits for the $\alpha$ and $\beta$ parameters.}
\end{table}
\endlandscape

\clearpage

\landscape
\begin{table}[htbp]
\centering
\label{table:entropyGC}
\begin{tabular}{cc}
& \\
\includegraphics{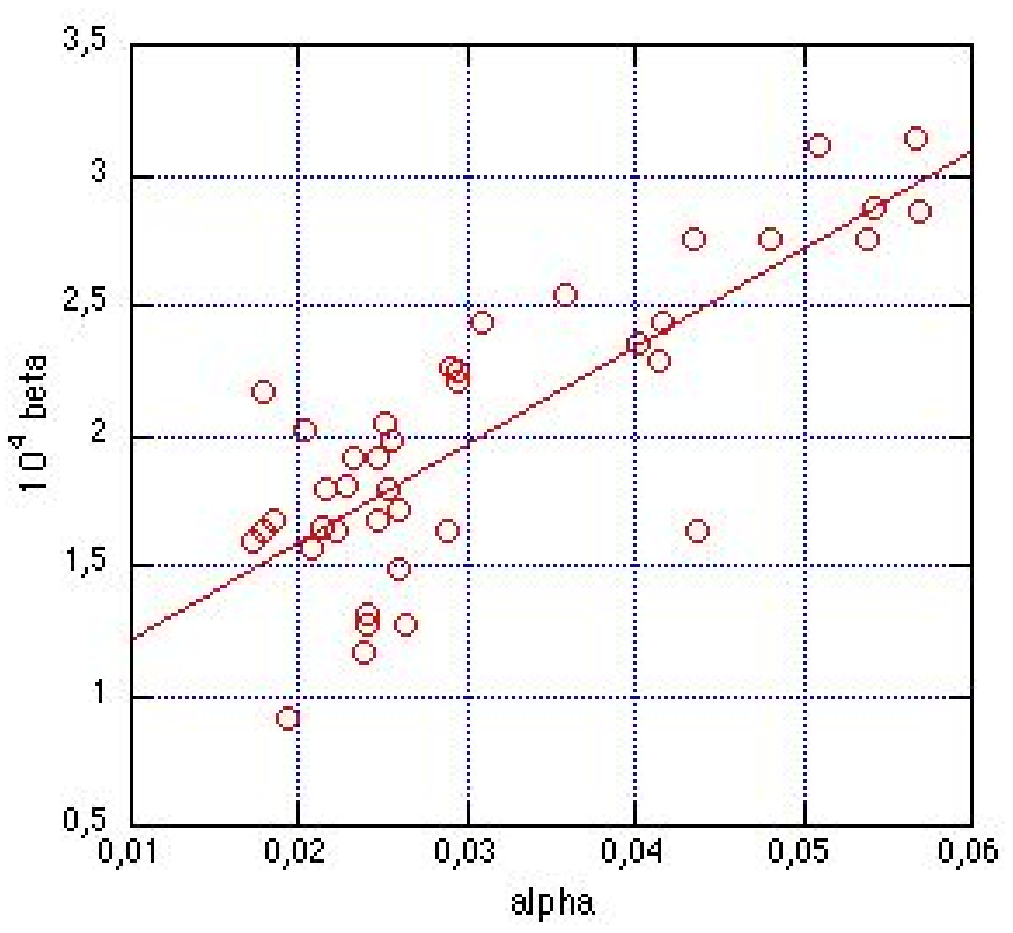} & \includegraphics{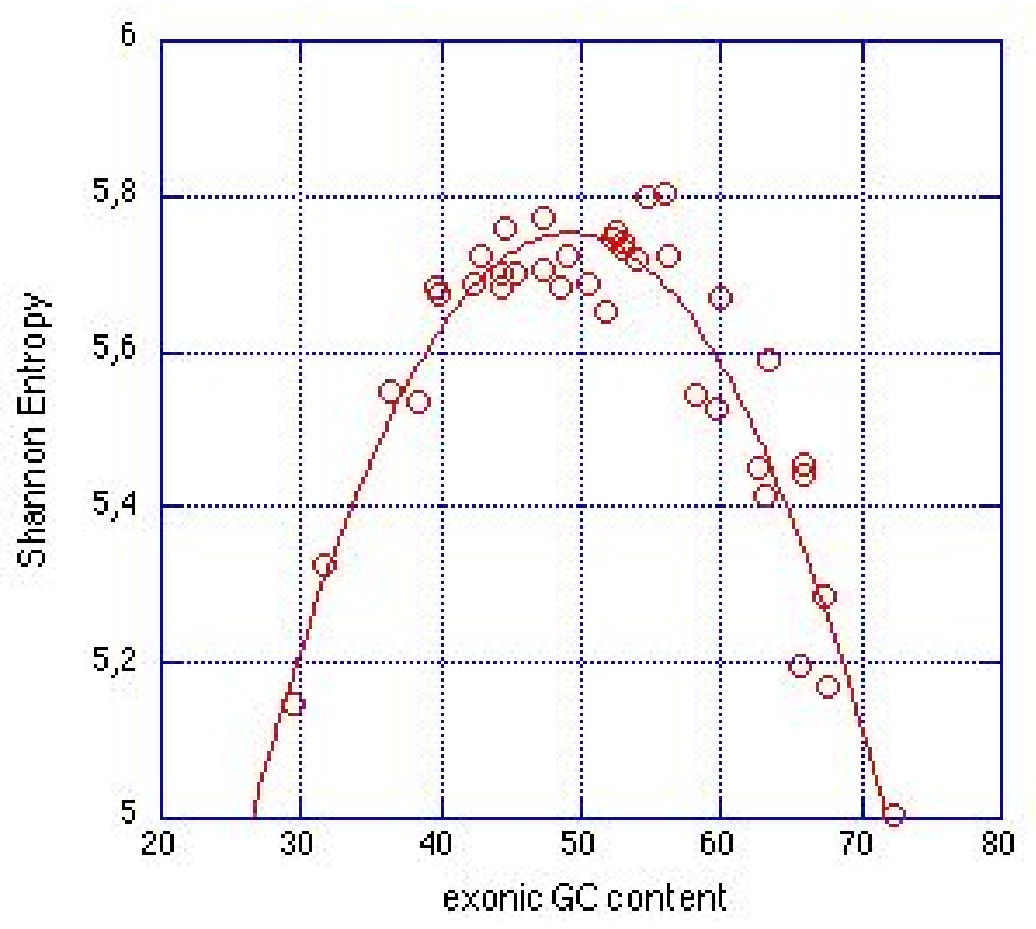} \\
& \\
Fit for $\beta$ as a function of $\alpha$ & Shannon entropy vs. the exonic 
$GC$ content \\
\end{tabular}
\caption{Fits for the $\alpha$, $\beta$ parameters and for the Shannon 
entropy.}
\end{table}
\endlandscape

\clearpage

\landscape
\begin{table}[htbp]
\scriptsize
\caption{Type of used codons of the observed rank distribution $f(n)$.
\label{table:codons}}
\begin{tabular}{|l|c|c|c|c|c|c|c|c|c|c|c|c|c|c|c|c|c|c|c|c|}
\hline
Rank & 1 & 2 & 3 & 4 & 5 & 6 & 7 & 8 & 9 & 10 & 11 & 12 & 13 & 14 & 15 & 
16 & 17 & 18 & 19 & 20 \\
\hline
Homo sapiens & GAG & CUG & CAG & AAG & GAA & GUG & GCC & GAC & AAA & GGC & 
AUG & GAU & AUC & UUC & CCC & AAC & CUC & AGC & ACC & GCU \\
Mus musculus & CUG & GAG & AAG & CAG & GUG & GAC & GAA & GCC & AUC & AUG & 
GGC & UUC & GAU & AAA & AAC & CUC & GCU & AGC & ACC & CCC \\
Rattus norvegicus & CUG & GAG & AAG & CAG & GUG & GAC & GCC & GAA & AUC & 
UUC & AUG & AAC & GGC & CUC & GAU & AAA & ACC & AGC & GCU & CCC \\
Gallus gallus & GAG & CUG & AAG & CAG & GAA & GUG & AAA & GAC & GCC & GAU & 
AAC & AUC & AUG & GGC & AGC & UUC & GCU & CCC & UAC & ACC \\
Xenopus laevis & GAA & GAG & AAA & AAG & CAG & GAU & CUG & AUG & GAC & AAU 
& AAC & GGA & GUG & GCU & CCA & AUU & GCA & UUU & UCU & ACA \\
Bos taurus & CUG & GAG & AAG & CAG & GUG & GCC & GAC & GAA & AUC & GGC & 
UUC & AAC & AUG & AAA & ACC & GAU & CUC & CCC & UAC & AGC \\
Arabidopsis thaliana & GAU & GAA & AAG & GAG & AAA & GCU & GUU & UCU & AUG 
& CUU & GGA & AAU & GGU & UUU & AUU & UUG & AAC & UUC & CAA & AGA \\
Oryza sativa japonica & GAG & GCC & GGC & AAG & GAC & GCG & CUC & GUG & GAU 
& AUG & UUC & CUG & GAA & CAG & GUC & AUC & CCG & GCU & AAC & CGC \\
Oryza sativa & GAG & AAG & GCC & GGC & GAC & GAU & CUC & AUG & GUG & GCG & 
UUC & CAG & GAA & AUC & GCU & AAC & GUC & CUG & GCA & GGG \\
Neurospora crassa & GAG & AAG & GCC & GAC & GGC & AAC & CUC & AUC & CAG & 
GUC & ACC & GAU & CCC & UUC & AUG & GAA & GCU & UCC & GGU & UAC \\
Drosophila melanogaster & GAG & AAG & CUG & CAG & GCC & GUG & GAU & GGC & 
AAC & GAC & AUG & AUC & UUC & ACC & GAA & AAU & AGC & UCC & UAC & CGC \\
Caenorhabditis elegans & GAA & AAA & GAU & AUU & GGA & AAU & CAA & AUG & 
AAG & CCA & UUU & UUC & GAG & GUU & GCU & CUU & UCA & UUG & ACA & GCA \\
Leishmania major & GCG & GAG & GCC & CUG & GUG & GGC & GAC & CAG & CGC & 
AAG & CCG & AGC & CUC & ACG & AUG & AAC & UCG & CAC & GCA & UAC \\
Sacch. cerevisiae & GAA & AAA & GAU & AAU & AAG & AUU & CAA & UUG & UUA & 
UUU & AAC & GGU & UCU & GUU & AGA & GCU & AUG & GAC & ACU & GAG \\
Schizosacch. pombe & GAA & AAA & GAU & AUU & AAU & UUU & UCU & GCU & GUU & 
CAA & UUA & CUU & AAG & UUG & ACU & UAU & CCU & GGU & GAG & AUG \\
Escherichia coli & CUG & GAA & AAA & GAU & GCG & AUU & CAG & GGC & AUG & 
GGU & GUG & GCC & AUC & UUU & ACC & AAC & GCA & CCG & AAU & CGU \\
Bacillus subtilis & AAA & GAA & AUU & GAU & UUU & AUC & AUG & GGC & GAG & 
CUG & CUU & UAU & AAU & ACA & GGA & GCA & AAG & GCG & CAA & UUA \\
Pseudom. aeruginosa & CUG & GCC & GGC & CGC & GAC & GCG & AUC & GAG & CAG & 
GUG & UUC & ACC & CCG & GUC & CUC & AAG & AGC & GAA & AAC & AUG \\
Mesorhizobium loti & GGC & GCC & CUG & AUC & GCG & GUC & GAC & CGC & UUC & 
GAG & CCG & AAG & CUC & GUG & ACC & CAG & AUG & GAA & UCG & GAU \\
Streptom. coelicolor A3 & GCC & GGC & CUG & GAC & GCG & GAG & GUC & ACC & 
CGC & CUC & GUG & CCG & CGG & AUC & UUC & CCC & CAG & CAC & UCC & AAG \\
Sinorhizobium meliloti & GGC & GCC & GCG & AUC & GUC & CUC & CUG & GAC & 
CGC & GAG & UUC & CCG & AAG & GAA & AUG & CAG & GUG & ACC & ACG & UCG \\
Nostoc sp. PCC 7120 & GAA & AUU & CAA & UUA & AAA & GAU & AAU & UUU & GCU & 
GGU & GCA & UUG & GUU & ACU & UAU & GUA & ACA & AUC & GCC & AUG \\
Agrobact. tumefaciens & GCC & GGC & CUG & AUC & GCG & GAA & CGC & GUC & UUC 
& GAU & GAC & AAG & CUC & CCG & AUG & GUG & CAG & GAG & ACC & ACG \\
Ralstonia solanacearum & CUG & GCC & GGC & GCG & CGC & GUG & AUC & GAC & 
CCG & CAG & GAG & UUC & ACC & GUC & AAG & ACG & AUG & AAC & UCG & CUC \\
Yersinia pestis & CUG & GAU & GAA & AAA & AUU & GCC & AUG & GGU & CAG & AAU 
& GCG & GGC & CAA & AUC & UUG & GUG & UUU & ACC & UUA & GAG \\
Methanosarc. acetivorans & GAA & AAA & CUU & GAU & GGA & AUU & GCA & AUC & 
GAG & UUU & CUG & GAC & AUG & AAU & AAG & AAC & AUA & GUU & UAU & UUC \\
Vibrio cholerae & GAA & GAU & AAA & CAA & AUU & GCG & GUG & UUU & CUG & GGU 
& AUG & AUC & GAG & GGC & UUG & AAU & GCC & UUA & GCU & ACC \\
Escherichia coli K12 & CUG & GAA & GCG & AAA & GAU & AUU & GGC & CAG & AUG 
& GUG & GCC & AUC & GGU & ACC & CCG & UUU & CGC & AAC & CGU & GCA \\
Mycobact. tuber. CDC1551 & GCC & CUG & GGC & GCG & GAC & GUG & ACC & AUC & 
GUC & CCG & GAG & CGC & CGG & CAG & UUC & UCG & AAC & GGG & GGU & AUG \\
Mycobact. tuber. H37Rv & GCC & GGC & CUG & GCG & GAC & GUG & ACC & AUC & 
GUC & CCG & GAG & CGC & CGG & UUC & CAG & AAC & UCG & GGG & GGU & AUG \\
Bacillus halodurans & GAA & AUU & AAA & GAU & UUU & GAG & CAA & UUA & AUG & 
AUC & UAU & CUU & GGA & GUU & AAG & ACG & AAU & GCA & GUG & GCG \\
Clostridium acetobutylicum & AAA & AUA & AAU & GAA & GAU & UUU & UUA & AUU 
& UAU & GGA & AAG & GUU & GUA & CUU & GCA & AUG & AGA & GCU & ACA & GGU \\
Caulobacter crescentus CB15 & GCC & CUG & GGC & GCG & GAC & CGC & AUC & GUC 
& GAG & ACC & AAG & UUC & GUG & CCG & CAG & AUG & UCG & AAC & CCC & CUC \\
Synechocystis sp. PCC 6803 & GAA & AUU & GCC & CAA & GAU & UUG & AAA & UUU 
& GUG & ACC & UUA & CCC & AAU & GGC & CAG & CUG & GCU & GGU & AUG & GAC \\
Sulfolobus solfataricus & AUA & UUA & AAA & GAA & AAG & GAU & AUU & AAU & 
UAU & GAG & GUA & GUU & UUU & GGA & AGA & GCU & GGU & AUG & ACU & GCA \\
Mycobacterium leprae & GCC & CUG & GUG & GAC & GCG & GGC & AUC & GUC & ACC 
& GAG & CCG & UUG & GGU & CGC & CAG & GAU & GAA & GCU & UUC & CGG \\
Brucella melitensis & GGC & GCC & CUG & GAA & CGC & GCG & AUC & GAU & AAG & 
GUG & CCG & UUC & AUG & CAG & CUU & GAC & GUC & ACC & CUC & GAG \\
Deinococcus radiodurans & CUG & GCC & GGC & GUG & GCG & GAC & CGC & ACC & 
CAG & CUC & GAG & CCC & GAA & CCG & AGC & GUC & AUC & UUC & GGG & CGG \\
Listeria monocytogenes & AAA & GAA & AUU & GAU & UUA & AAU & UUU & CAA & 
GCA & GUU & AUG & ACA & GGU & UAU & GCU & GUA & CUU & GGA & AUC & CCA \\
Clostridium perfringens & AAA & GAA & UUA & AUA & AAU & GAU & GGA & UUU & 
GUU & UAU & AUU & GCU & AGA & AAG & GUA & AUG & ACU & UCA & GCA & ACA \\
\hline
\end{tabular}
\end{table}
\endlandscape

\clearpage

\begin{table}[htbp]
\centering
\caption{Values of the best-fit parameters for the amino-acids.
\label{table:amino}}
\begin{tabular}{|l|c|c|c|}
\hline
Species & $10^{3}\,B$ & $F_{0}$ & $\chi^2$ 
\\
\hline
\hline
Homo sapiens & 3.8 & 0.089 & 0.0072 \\
\hline
Arabidopsis thaliana & 3.8 & 0.090 & 0.0068 \\
\hline
Drosophila melanogaster & 3.5 & 0.087 & 0.0125 \\
\hline
Caenorhabditis elegans & 3.3 & 0.084 & 0.0124 \\
\hline
Mus musculus & 3.7 & 0.088 & 0.0087 \\
\hline
Saccharomyces cerevisiae & 3.9 & 0.090 & 0.0121 \\
\hline
Escherichia coli & 4.0 & 0.091 & 0.0115 \\
\hline
Rattus norvegicus & 3.7 & 0.088 & 0.0084 \\
\hline
Oryza sativa japonica & 4.1 & 0.093 & 0.0057 \\
\hline
Schizosaccharomyces pombe & 3.8 & 0.089 & 0.0162 \\
\hline
Bacillus subtilis & 4.0 & 0.091 & 0.0104 \\
\hline
Pseudomonas aeruginosa & 4.9 & 0.101 & 0.0493 \\
\hline
Mesorhizobium loti & 4.7 & 0.100 & 0.0215 \\
\hline
Streptomyces coelicolor A3 & 5.6 & 0.109 & 0.0624 \\
\hline
Sinorhizobium meliloti & 4.7 & 0.100 & 0.0188 \\
\hline
Nostoc sp. PCC 7120 & 4.0 & 0.092 & 0.0174 \\
\hline
Oryza sativa & 3.9 & 0.091 & 0.0028 \\
\hline
Agrobacterium tumefaciens str. C58 & 4.6 & 0.098 & 0.0144 \\
\hline
Ralstonia solanacearum & 4.7 & 0.101 & 0.0351 \\
\hline
Yersinia pestis & 4.0 & 0.092 & 0.0135 \\
\hline
Methanosarcina acetivorans str. C2A & 4.1 & 0.092 & 0.0063 \\
\hline
Vibrio cholerae & 3.9 & 0.091 & 0.0148 \\
\hline
Escherichia coli K12 & 4.0 & 0.091 & 0.0154 \\
\hline
Mycobacterium tuberculosis CDC1551 & 5.2 & 0.105 & 0.01121 \\
\hline
Mycobacterium tuberculosis H37Rv & 5.3 & 0.106 & -- \\
\hline
Bacillus halodurans & 4.0 & 0.091 & 0.0100 \\
\hline
Clostridium acetobutylicum & 4.6 & 0.097 & 0.0076 \\
\hline
Caulobacter crescentus CB15 & 5.1 & 0.104 & 0.0524 \\
\hline
Gallus gallus & 3.6 & 0.088 & 0.0040 \\
\hline
Synechocystis sp. PCC 6803 & 4.1 & 0.093 & 0.0168 \\
\hline
Sulfolobus solfataricus & 4.4 & 0.096 & 0.0143 \\
\hline
Mycobacterium leprae & 4.9 & 0.101 & 0.0401 \\
\hline
Brucella melitensis & 4.5 & 0.097 & 0.0142 \\
\hline
Deinococcus radiodurans & 5.2 & 0.105 & 0.0679 \\
\hline
Xenopus laevis & 3.5 & 0.086 & 0.0084 \\
\hline
Listeria monocytogenes & 4.2 & 0.093 & 0.0088 \\
\hline
Neurospora crassa & 4.0 & 0.091 & 0.0042 \\
\hline
Clostridium perfringens & 4.6 & 0.098 & 0.0035 \\
\hline
Leishmania major & 4.7 & 0.099 & 0.0367 \\
\hline
Bos taurus & 3.6 & 0.087 & 0.0082 \\
\hline
\end{tabular}
\end{table}

\clearpage

\landscape
\begin{table}[htbp]
\centering
\caption{Amino-acids rank distributions.
\label{table:figures}}
\begin{tabular}{cc}
\includegraphics{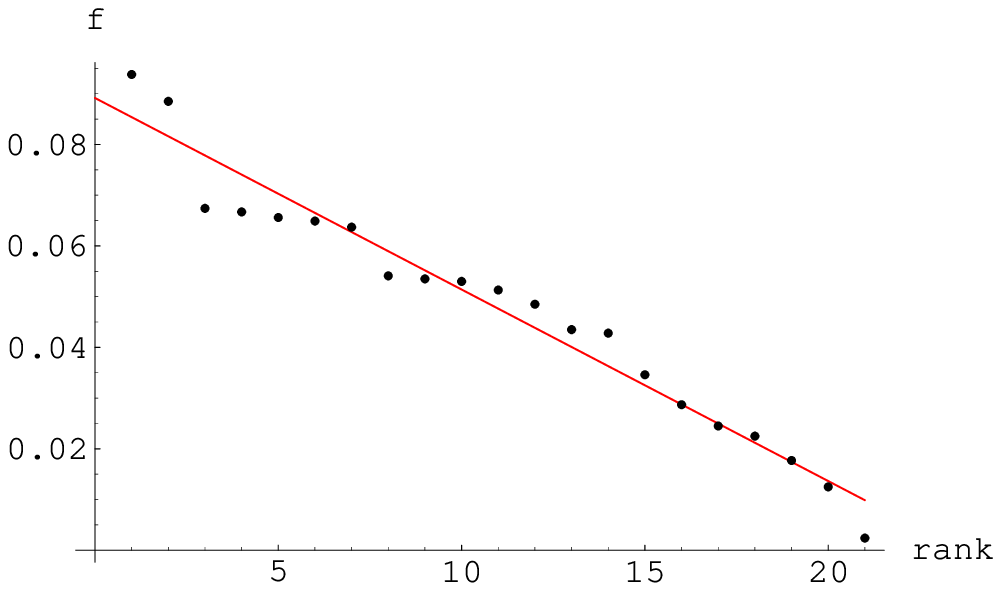} & \includegraphics{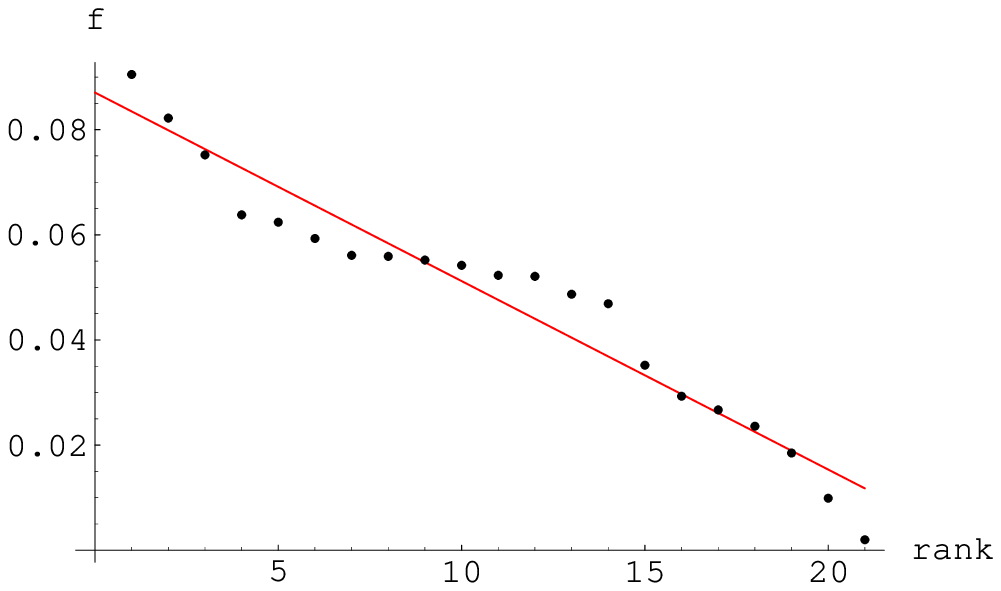} \\
Arabidopsis thaliana & Drosophila melanogaster \\
& \\
\includegraphics{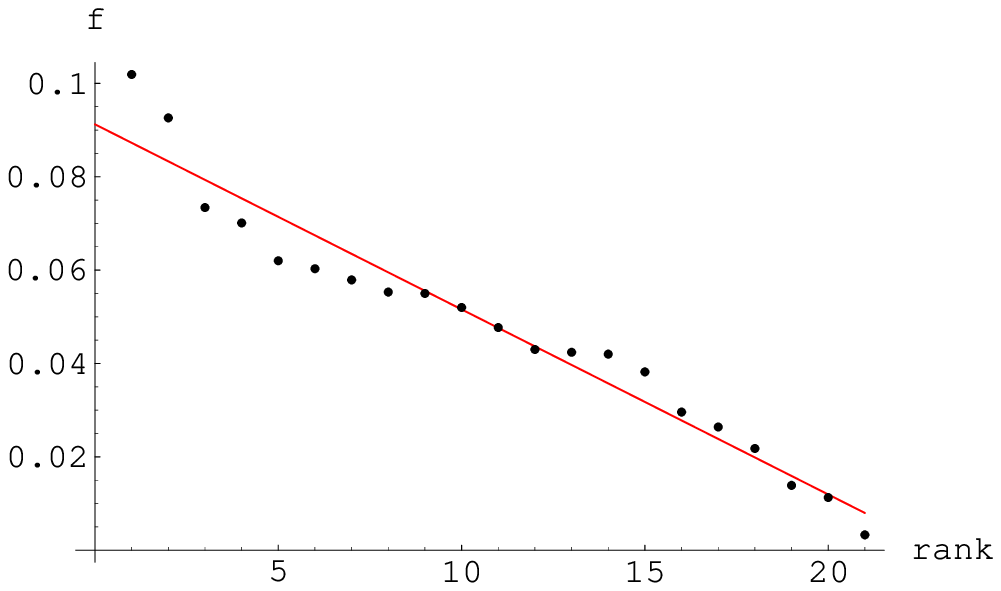} & \includegraphics{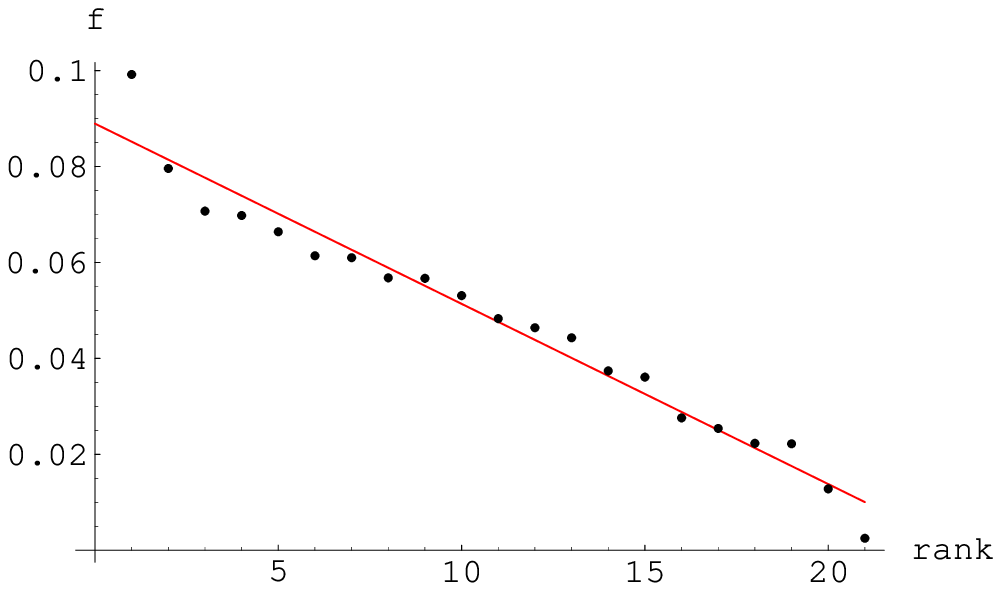} \\
Escherichia coli & Homo sapiens \\
\end{tabular}
\end{table}
\endlandscape

\clearpage

\landscape
\begin{table}[htbp]
\centering
\caption{Amino-acids rank distributions : linear vs. cubic fits.
\label{table:lincub}}
\begin{tabular}{cc}
& \\
\includegraphics{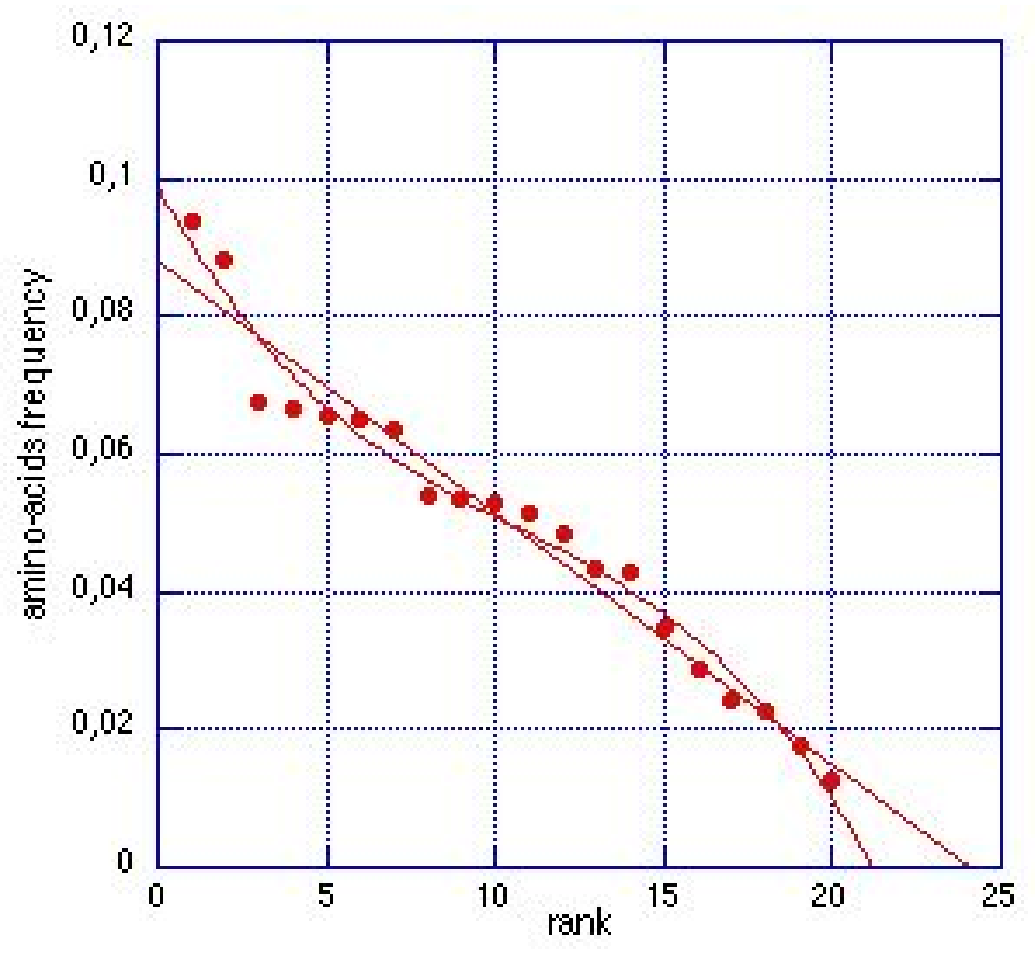} & \includegraphics{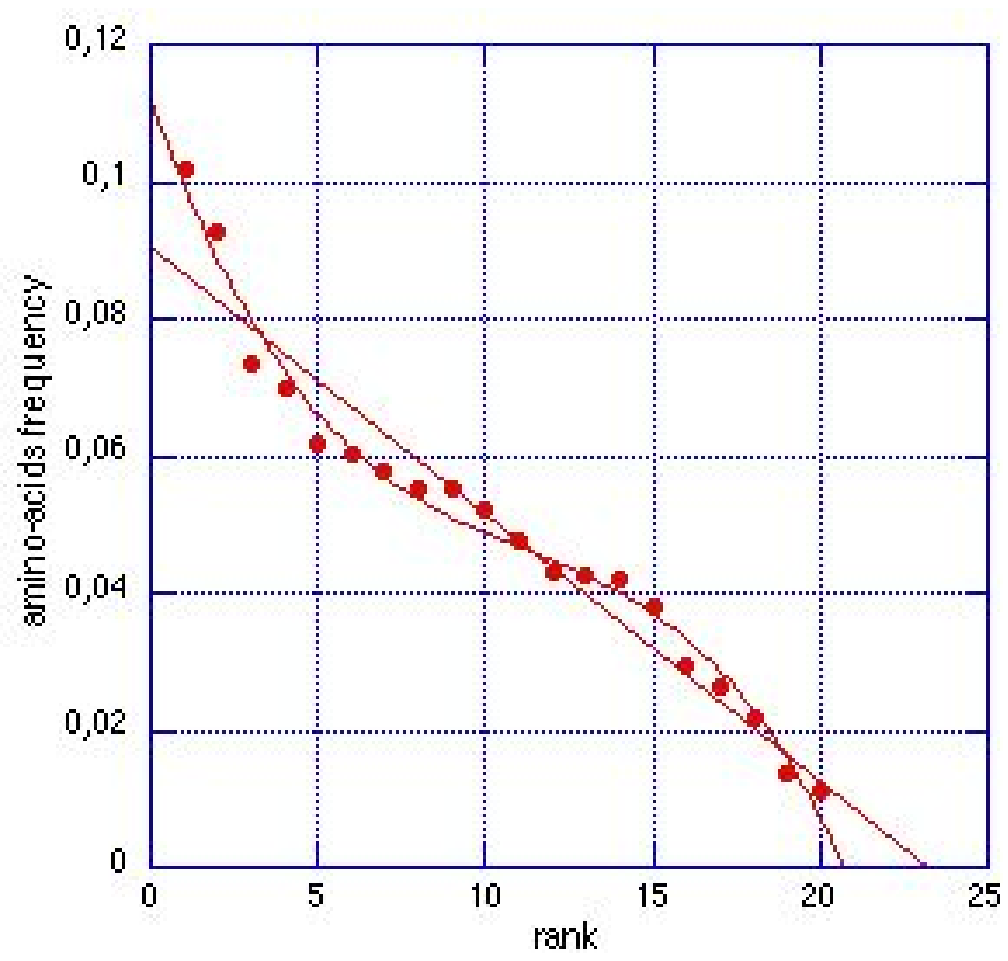} \\
& \\
Arabidopsis thaliana & Escherichia coli \\
\end{tabular}
\end{table}
\endlandscape

\clearpage

\landscape
\begin{table}[htbp]
\scriptsize \caption{Type of used amino-acids of the observed rank 
distribution.
\label{table:acidamines}}
\begin{tabular}{|l|c|c|c|c|c|c|c|c|c|c|c|c|c|c|c|c|c|c|c|c|}
\hline
Rank & 1 & 2 & 3 & 4 & 5 & 6 & 7 & 8 & 9 & 10 & 11 & 12 & 13 & 14 & 15 & 
16 & 17 & 18 & 19 & 20 \\
\hline
Homo sapiens & Leu & Ser & Ala & Glu & Gly & Val & Pro & Lys & Arg & Thr & 
Asp & Gln & Ile & Phe & Asn & Tyr & His & Met & Cys & Trp \\
Mus musculus & Leu & Ser & Ala & Gly & Glu & Val & Pro & Lys & Arg & Thr & 
Asp & Ile & Gln & Phe & Asn & Tyr & His & Cys & Met & Trp \\
Rattus norvegicus & Leu & Ser & Ala & Glu & Gly & Val & Pro & Lys & Thr & 
Arg & Asp & Ile & Gln & Phe & Asn & Tyr & His & Met & Cys & Trp \\
Gallus gallus & Leu & Ser & Glu & Ala & Gly & Lys & Val & Pro & Thr & Arg & 
Asp & Ile & Gln & Asn & Phe & Tyr & His & Met & Cys & Trp \\
Xenopus laevis & Leu & Ser & Glu & Lys & Ala & Gly & Val & Pro & Thr & Asp 
& Arg & Ile & Gln & Asn & Phe & Tyr & Met & His & Cys & Trp \\
Bos taurus & Leu & Ser & Ala & Gly & Glu & Val & Lys & Pro & Thr & Arg & 
Asp & Ile & Gln & Phe & Asn & Tyr & Cys & His & Met & Trp \\
Arabidopsis thaliana & Leu & Ser & Val & Glu & Gly & Ala & Lys & Asp & Arg 
& Ile & Thr & Pro & Asn & Phe & Gln & Tyr & Met & His & Cys & Trp \\
Oryza sativa japonica & Ala & Leu & Gly & Ser & Arg & Val & Glu & Pro & Asp 
& Thr & Lys & Ile & Phe & Gln & Asn & His & Tyr & Met & Cys & Trp \\
Oryza sativa & Ala & Leu & Gly & Ser & Arg & Val & Glu & Pro & Asp & Lys & 
Thr & Ile & Phe & Gln & Asn & Tyr & His & Met & Cys & Trp \\
Neurospora crassa & Ala & Leu & Ser & Gly & Glu & Pro & Arg & Thr & Val & 
Asp & Lys & Ile & Gln & Asn & Phe & Tyr & His & Met & Trp & Cys \\
Drosophila melanogaster & Leu & Ser & Ala & Glu & Gly & Val & Lys & Thr & 
Arg & Pro & Asp & Gln & Ile & Asn & Phe & Tyr & His & Met & Cys & Trp \\
Caenorhabditis elegans & Leu & Ser & Glu & Lys & Ala & Val & Ile & Thr & 
Gly & Arg & Asp & Asn & Phe & Pro & Gln & Tyr & Met & His & Cys & Trp \\
Leishmania major & Ala & Leu & Ser & Arg & Val & Gly & Thr & Pro & Glu & 
Asp & Gln & Lys & Ile & His & Phe & Asn & Tyr & Met & Cys & Trp \\
Sacch. cerevisiae & Leu & Ser & Lys & Ile & Glu & Asn & Thr & Asp & Val & 
Ala & Gly & Arg & Phe & Pro & Gln & Tyr & His & Met & Cys & Trp \\
Schizosacch. pombe & Leu & Ser & Glu & Lys & Ala & Ile & Val & Thr & Asp & 
Asn & Gly & Arg & Pro & Phe & Gln & Tyr & His & Met & Cys & Trp \\
Escherichia coli & Leu & Ala & Gly & Val & Ser & Ile & Glu & Thr & Arg & 
Asp & Lys & Gln & Pro & Asn & Phe & Tyr & Met & His & Trp & Cys \\
Bacillus subtilis & Leu & Ala & Ile & Glu & Lys & Gly & Val & Ser & Thr & 
Asp & Phe & Arg & Asn & Gln & Pro & Tyr & Met & His & Trp & Cys \\
Pseudom. aeruginosa & Leu & Ala & Gly & Arg & Val & Glu & Ser & Asp & Pro & 
Gln & Thr & Ile & Phe & Lys & Asn & Tyr & His & Met & Trp & Cys \\
Mesorhizobium loti & Ala & Leu & Gly & Val & Arg & Ser & Asp & Ile & Glu & 
Thr & Pro & Phe & Lys & Gln & Asn & Met & Tyr & His & Trp & Cys \\
Streptom. coelicolor A3 & Ala & Leu & Gly & Val & Arg & Pro & Thr & Asp & 
Glu & Ser & Ile & Phe & Gln & His & Lys & Tyr & Asn & Met & Trp & Cys \\
Sinorhizobium meliloti & Ala & Leu & Gly & Val & Arg & Glu & Ser & Ile & 
Asp & Thr & Pro & Phe & Lys & Gln & Asn & Met & Tyr & His & Trp & Cys \\
Nostoc sp. PCC 7120 & Leu & Ala & Ile & Val & Gly & Ser & Glu & Thr & Gln & 
Arg & Lys & Asp & Pro & Asn & Phe & Tyr & His & Met & Trp & Cys \\
Agrobact. tumefaciens & Ala & Leu & Gly & Val & Arg & Ser & Glu & Ile & Asp 
& Thr & Pro & Phe & Lys & Gln & Asn & Met & Tyr & His & Trp & Cys \\
Ralstonia solanacearum & Ala & Leu & Gly & Val & Arg & Thr & Asp & Pro & 
Ser & Glu & Ile & Gln & Phe & Lys & Asn & Tyr & His & Met & Trp & Cys \\
Yersinia pestis & Leu & Ala & Gly & Val & Ser & Ile & Glu & Thr & Arg & Asp 
& Gln & Lys & Pro & Asn & Phe & Tyr & Met & His & Trp & Cys \\
Methanosarc. acetivorans & Leu & Glu & Ile & Gly & Ser & Ala & Val & Lys & 
Thr & Asp & Arg & Asn & Phe & Pro & Tyr & Gln & Met & His & Cys & Trp \\
Vibrio cholerae & Leu & Ala & Val & Gly & Ser & Ile & Glu & Thr & Asp & Gln 
& Lys & Arg & Phe & Asn & Pro & Tyr & Met & His & Trp & Cys \\
Escherichia coli K12 & Leu & Ala & Gly & Val & Ile & Ser & Glu & Arg & Thr 
& Asp & Gln & Pro & Lys & Asn & Phe & Tyr & Met & His & Trp & Cys \\
Mycobact. tuber. CDC1551 & Ala & Leu & Gly & Val & Arg & Thr & Pro & Asp & 
Ser & Glu & Ile & Gln & Phe & Asn & His & Tyr & Lys & Met & Trp & Cys \\
Mycobact. tuber. H37Rv & Ala & Gly & Leu & Val & Arg & Thr & Asp & Pro & 
Ser & Glu & Ile & Gln & Phe & Asn & His & Tyr & Lys & Met & Trp & Cys \\
Bacillus halodurans & Leu & Glu & Val & Ala & Gly & Ile & Lys & Ser & Thr & 
Asp & Arg & Phe & Gln & Pro & Asn & Tyr & Met & His & Trp & Cys \\
Clostridium acetobutylicum & Ile & Lys & Leu & Ser & Glu & Val & Asn & Gly 
& Ala & Asp & Thr & Phe & Tyr & Arg & Pro & Met & Gln & His & Cys & Trp \\
Caulobacter crescentus CB15 & Ala & Leu & Gly & Val & Arg & Asp & Pro & Glu 
& Thr & Ser & Ile & Phe & Lys & Gln & Asn & Met & Tyr & His & Trp & Cys \\
Synechocystis sp. PCC 6803 & Leu & Ala & Gly & Val & Ile & Glu & Ser & Gln 
& Thr & Pro & Arg & Asp & Lys & Asn & Phe & Tyr & Met & His & Trp & Cys \\
Sulfolobus solfataricus & Leu & Ile & Lys & Val & Glu & Ser & Gly & Ala & 
Asn & Tyr & Arg & Thr & Asp & Phe & Pro & Gln & Met & His & Trp & Cys \\
Mycobacterium leprae & Ala & Leu & Val & Gly & Arg & Thr & Ser & Asp & Pro 
& Glu & Ile & Gln & Phe & Lys & Asn & His & Tyr & Met & Trp & Cys \\
Brucella melitensis & Ala & Leu & Gly & Val & Arg & Ile & Glu & Ser & Asp & 
Thr & Pro & Lys & Phe & Gln & Asn & Met & Tyr & His & Trp & Cys \\
Deinococcus radiodurans & Ala & Leu & Gly & Val & Arg & Pro & Thr & Glu & 
Ser & Asp & Gln & Ile & Phe & Lys & Asn & Tyr & His & Met & Trp & Cys \\
Listeria monocytogenes & Leu & Ile & Ala & Glu & Lys & Val & Gly & Thr & 
Ser & Asp & Asn & Phe & Arg & Pro & Gln & Tyr & Met & His & Trp & Cys \\
Clostridium perfringens & Ile & Leu & Lys & Glu & Gly & Val & Asn & Ser & 
Asp & Ala & Thr & Phe & Tyr & Arg & Pro & Met & Gln & His & Cys & Trp \\
\hline
\end{tabular}
\end{table}
\endlandscape

\end{document}